\documentclass[letterpaper,aps,prd,twocolumn,tightenlines,preprintnumbers,nofootinbib,showkeys,superscriptaddress,longbibliography]{revtex4-1}

\usepackage{lipsum} 
\usepackage{booktabs}
\usepackage{XCharter}
\usepackage[T1]{fontenc}
\usepackage{bm}
\usepackage{bbold}
\usepackage{pifont}
\usepackage{rotating}
\usepackage{fullpage}
\usepackage{amsfonts,shorthand}
\usepackage{amsmath}
\usepackage{slashed}
\usepackage{siunitx}
\usepackage{amssymb}
\usepackage{graphicx}
\usepackage{epic}
\usepackage{eepic}
\usepackage{epsfig}
\usepackage{latexsym}
\usepackage[dvipsnames]{xcolor}
\usepackage[export]{adjustbox}
\usepackage{float}
\usepackage{multirow, makecell}
\usepackage{hyperref}
\usepackage{enumitem}
\hypersetup{colorlinks=true,citecolor=blue,linkcolor=NavyBlue,urlcolor=red}
\usepackage[caption=false]{subfig}
 
\usepackage{natbib}
\usepackage{relsize}
\usepackage[left=1.65cm,right=1.65cm,top=1.83cm,bottom=1.84cm]{geometry}
\newcommand{\orcid}[1]{\hspace{1mm}\href{https://orcid.org/#1}{\includegraphics[height=0.3cm,keepaspectratio]{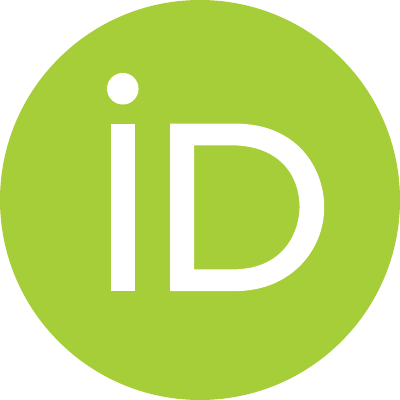}}}

\begin{document}
\relscale{1.05}
\title{A comprehensive study of the charged Higgs boson in the two Higgs doublet type-II seesaw model }

\author{Brahim Ait-Ouazghour\orcid{0009-0006-1419-969X}}
\email{b.ouazghour@gmail.com}
\affiliation{Cadi Ayyad University,  Faculty of Science Semlalia,  LPHEAG, P.O.B. 2390 Marrakech 40000, Morocco.}
\author{Abdesslam Arhrib\orcid{0000-0001-5619-7189}}
\email{aarhrib@gmail.com}
\affiliation{Abdelmalek Essaadi University, FST Tanger B.P. 416, Morocco}
\affiliation{Department of Physics and CTC, National Tsing Hua University, Hsinchu, Taiwan 300}
\author{Mohamed Chabab\orcid{0000-0002-2772-4290}}
\email{mchabab@uca.ac.ma}
\affiliation{Cadi Ayyad University,  Faculty of Science Semlalia,  LPHEAG, P.O.B. 2390 Marrakech 40000, Morocco.}
\affiliation{Cadi Ayyad University, National School of Applied Science, P.O.B. 63 Safi 46000, Morocco}
\author{Khalid Goure\orcid{0009-0007-5292-5012}}
\email{khalidgoure01@gmail.com}
\affiliation{Cadi Ayyad University,  Faculty of Science Semlalia,  LPHEAG, P.O.B. 2390 Marrakech 40000, Morocco.}

\date{\today}

\begin{abstract} 
We investigate the phenomenology of the charged Higgs boson within the Two-Higgs-Doublet Type-II Seesaw Model (2HDMcT Type-III) at future $\mu^+ \mu^-$ colliders. Focusing on $2 \to 2$ processes such as $\mu^+ \mu^- \to H^+_1 S^-$ ($S^- = H^-_1, H^-_2$) and $\mu^+ \mu^- \to H^+_1 W^-$, we incorporate both theoretical and experimental constraints to assess their production prospects. We find that $\sigma(\mu^+ \mu^- \to H_1^+ H_1^-)$ can reach or exceed the corresponding $e^+ e^-$ cross section, while $\sigma(\mu^+ \mu^- \to H_1^+ H_2^-)$ is roughly eighteen times smaller but can be enhanced up to 0.62 fb upon relaxing electroweak precision observables constraints. Moreover, we observe that the cross section for $\mu^+ \mu^- \to H^+_1 W^-$ is significantly enhanced due to the large $\tan\beta$ amplification characteristic of the Type-III scenario. Furthermore, we conduct a signal-background analysis and determine the discovery ($5\sigma$) regions at a 3 TeV muon collider for the $\mu^+ \mu^- \to H_1^+ H_1^-$, $\mu^+ \mu^- \to H_1^+ H_2^-$, and $\mu^+ \mu^- \to W^\pm H_1^\mp$ processes.
\end{abstract}
\maketitle

\section{Introduction}

The Standard Model (SM) of particle physics has been rigorously tested with remarkable precision, but although it currently represent our most advanced understanding of particle physics phenomenology, there are several indications that the SM is only the low-energy approximation of a more fundamental theory. Among these drawbacks facing the SM, we can cite the origin of dark matter\cite{Zwicky:1933gu,Rubin:1970zza}, the hierarchy problem \cite{Veltman:1980mj},  and the neutrino mass generation \cite{RevModPhys.88.030501,RevModPhys.88.030502}. 

To address these problems, several attempt were made to go Beyond the Standard Model (BSM). Since the fermion sector is not of minimal form and given that there is no known fundamental reason restricting the scalar sector to be minimal. Following this reasoning, many extensions of the SM have explored extended Higgs sectors, as the popular two Higgs Doublet Model (2HDM) \cite{Deshpande:1977rw,PhysRevD.15.1958,Branco:2011iw,Dawson:2018dcd,Ivanov:2017dad,PhysRevLett.43.1566}, the Higgs Triplet Models (HTM) \cite{PhysRevLett.43.1566,Dawson:2018dcd} and recently 2HDM augmented by a complex triplet scalar, dubbed the two Higgs Doublet Type-II Seesaw model (2HDMcT) \cite{Ouazghour:2018mld,Ouazghour:2023eqr,Ouazghour:2024fgo,BrahimAit-Ouazghour:2024img}.

Since the  mass generation from seesaw mechanism is similar to the Brout-Englert-Higgs mechanism, 2HDMcT model is appealing, displaying many phenomenological features especially different from those emerging in 2HDM scalar sector. Apart its broader spectrum than 2HDM's one, the doubly charged Higgs $H^{++}$, as a smoking gun of 2HDMcT, $H^{++}$ is currently intensively searched for at ATLAS and CMS, by means of promising decay channels, such as $H^{++}H^{--}$ and $H_i^{+}H_i^{-}$ $(i=1,2)$, decaying to the same sign di-lepton \cite{Ouazghour:2018mld}. Furthermore, 2HDMcT is arguably one of the simplest frameworks that can account for both the dark matter issue \cite{Chen:2014lla} as well as the neutrino mass problem \cite{FileviezPerez:2008jbu,Cai:2017mow,King:2015aea}. Along side the doubly charged Higgs the phenomenological features of this model include the possibility of enhanced Higgs couplings, modified Higgs decay channels, and the presence of additional scalar particles that can be probed at the Large Hadron Collider (LHC) or other future colliders. For example one of the prospective signals that can be probed is $pp \to Z/\gamma \to H_2^+ H_2^- \to H^{++} W^- H^{--} W^+ \to l^+l^+l^-l^-+4j$. This process does not show up neither in 2HDM nor  in HTM giving that the mass splitting $\Delta m = m_{H^{++}} -  m_{H^+}$ is constrained by the oblique parameters to be less than  $40$ GeV in HTM \cite{Ashanujjaman:2021txz}. New contribution to the oblique parameters from the new states in 2HDMcT makes this avoidable. Additionally, beyond Higgs phenomenology, it has been demonstrated that interactions between doublet and triplet fields may induce a strong first order electroweak phase transition, which provide conditions for the generation of the baryon asymmetry through electroweak baryongenesis \cite{Ramsey-Musolf:2019lsf}. In view of this, adding an $SU(2)_{L}$ Higgs doublet and triplet fields results in a total of eleven physical Higgs particles \cite{Ouazghour:2018mld}.

One of the most intriguing features of models incorporating additional Higgs doublets, triplets, or higher representations is the presence of charged scalars, particles that carry electromagnetic charge. Detecting those would be a clear signal of physics BSM. The active pursuit of these new charged scalars has been a central focus and a major driving force behind current and future experiments, particularly at the LHC \cite{ATLAS:2017eiz,ATLAS:2017ayi,CMS:2018amk,ATLAS:2018oht,CMS:2019pzc,CMS:2019mij,2022arXiv220701046C}. Moreover, with the upcoming LHC run and its anticipated High Luminosity upgrade, we are entering an era of precise measurement programs to rigorously examine the SM \cite{BejarAlonso:2015ldp}. However, despite extensive searches, no convincing evidence has been found yet for the existence of new particles. Thus, these searches have only yielded upper limits on their potential signal cross section and can lead to significant constraints on the parameters of these BSM theories. At hadron colliders, the charged Higgs boson can be produced through several channels : a light charged Higgs boson can be produced through the top quark decay when the charged-Higgs mass is less than $m_t - m_b$. For a heavier charged Higgs boson, production can occur through associated production with a top quark through gluon-gluon fusion or gluon-bottom quark annihilation, and a number of other processes ( see Ref.\cite{Akeroyd:2016ymd}). At $e^+e^-$ colliders, the charged Higgs boson is predominantly produced via the $s$-channel process $e^+e^-\to \gamma^*, Z^*\to H^+H^-$ \cite{Komamiya:1988rs}. Consequently, the production rate is determined solely by the charged Higgs boson mass and the gauge couplings \cite{Arhrib:1998gr,Guasch:2001hk}. The contribution from $s$-channel diagrams involving neutral Higgs exchange is proportional to the electron mass ($m_e$) and is therefore significantly suppressed. Similarly, the associated production with a W boson $e^+ e^- \to W^\pm
H^\mp$ in the 2HDM \cite{Arhrib:1999rg,Kanemura:1999tg,Ouazghour:2023plc} and in the MSSM \cite{Heinemeyer:2016wey,Brein:2006xda,Logan:2002it} is only mediated at the one-loop level, with the tree-level cross section being suppressed by the electron mass $m_e$ \cite{Arhrib:1999rg,Kanemura:1999tg, 	Heinemeyer:2016wey,Brein:2006xda, Logan:2002it}.

The search for non-standard Higgs bosons at colliders, although unsuccessful so far, has established the current bounds on the masses and couplings of these non-SM Higgs bosons. The High Luminosity LHC (HL-LHC) will enhance some of these measurements and may provide hints of new physics. However, to continue the precise measurement program initiated at the LHC, there is a consensus on the need to construct a clean electron-positron Higgs factory. \cite{Abe:2001grn,LinearColliderAmericanWorkingGroup:2001rzk,LinearColliderAmericanWorkingGroup:2001tzv}. In comparison to hadron colliders, these colliders, with their cleaner $e^+e^-$ background, can significantly enhance the precision of measurements beyond those achieved at the LHC \cite{LCCPhysicsWorkingGroup:2019fvj} and can allow for detailed studies of the SM-like Higgs boson properties and couplings \cite{Gupta:2013zza,Baglio:2016bop,Tian:2016qlk,Durig:2016jrs,Liu:2018peg} which can offer a promising opportunity for a potential discovery of new physics. 
Several $e^+e^-$ colliders projects are currently planned, including the Compact Linear Collider (CLIC) \cite{CLICPhysicsWorkingGroup:2004qvu,Aicheler:2012bya}, the Circular Electron Positron Collider (CEPC)
\cite{An:2018dwb}, the
International Linear Collider (ILC)
\cite{LCCPhysicsWorkingGroup:2019fvj,Moortgat-Pick:2015lbx,Bhattacharya:2023mjr}, and the Future
Circular Collider (FCC-ee)
\cite{FCC:2018evy,TLEPDesignStudyWorkingGroup:2013myl}. 
To achieve our goal of a significant increase in the center-of-mass energy in particle collisions, a paradigm shift from the traditional technologies of proton-proton and electron-positron colliders is necessary. The idea of colliding beams of positively and negatively charged muons was first proposed in the late 1960s. However, there have been renewed interests in muon colliders
operating at high energies in the range of multi-TeV
\cite{Delahaye:2019omf,Han:2020uid,Long:2020wfp} in recent years.
Muons can be accelerated and collided in rings without experiencing the significant synchrotron radiation losses that limit the performance of electron-positron colliders. This allows a muon collider to use the traditional and well-established accelerator technologies of superconducting high-field magnets and RF cavities. In contrast to protons, where the collision energy is dispersed among constituent quarks and gluons, muons are point-like particles that deliver their entire energy $\sqrt{s}$ directly to the collision. This enables them to probe much higher energy scales compared to protons colliding at the same beam energy which makes this collisions much more efficient \cite{Long:2020wfp,Capdevilla:2021rwo,Liu:2021jyc,Huang:2021nkl,Yin:2020afe,Buttazzo:2020eyl,Capdevilla:2020qel,Han:2020pif,Han:2020uak,Costantini:2020stv}. Moreover, several studies have proposed employing the muon collider for detecting electroweak dark matter \cite{Han:2020uak} and uncovering heavy particles from BSM physics \cite{Costantini:2020stv,Han:2021udl,Vignaroli:2023rxr}. The utilization of muon colliders will mark the inaugural application of unstable particles in a collider.

Pair production of charged Higgs bosons and their associated production with a $W$ gauge boson at the upcoming muon collider has been investigated within the framework of the Minimal Supersymmetric Standard Model (MSSM) \cite{Akeroyd:1999xf,Hashemi:2012nz}, 2HDM with CP-violation \cite{Akeroyd:2000zs}, and more recently, in the CP-conserving scenario \cite{Ouazghour:2023plc}.

In this study, we examine the production of the lighter charged Higgs boson $H_1^\pm$ at a prospective muon collider within the framework of 2HDMcT, specifically focusing on the Type-III Yukawa texture scenario. In Type-II and Type-III, the Yukawa couplings of the neutral scalars to the muons, \( h_i \mu^+ \mu^- \) and \( A_i \mu^+ \mu^- \), are approximately proportional to \( \tan\beta \), since both scale as \( 1/\cos\beta = \sqrt{1 + \tan^2\beta} \approx \tan\beta \). Conversely, the remaining Yukawa textures are inversely proportional to $\tan \beta$, leading to suppressed couplings for large $\tan \beta$. While Type-II couplings are proportional to $\tan \beta$, the value of $\tan \beta$ is typically restricted by experimental bounds to be small\footnote{Constraints implemented via HiggsTools Subpackage \texttt{HiggsBounds}.}. Therefore, we anticipate a significant enhancement for large $\tan \beta$ in the Type-III scenario where there isn't such restriction.

Specifically, we analyze the $2\to2$ processes\footnote{~See also Refs~\cite{Ouazghour:2025owf,Ouazghour:2024twx} for $2\to3$ processes.} : $\mu^+ \mu^- \to H_1^\pm H_1^\mp$, $\mu^+ \mu^- \to H_1^\pm H_2^\mp$ and $\mu^+ \mu^- \to W^\pm H_1^\mp$.
Given the broader spectrum of the 2HDMcT compared to the 2HDM, these two processes receive additional contributions from $s$-channel neutral-Higgs exchange and gauge boson exchange. These new contributions may enhance or suppress the cross section\footnote{Depending on the nature of interference.} with respect to the one reported recently in the case of 2HDM \cite{Ouazghour:2023plc}. Additionally, considering that the muon mass is approximately 207 times greater than the electron mass, the cross-section at the muon collider can be significantly enhanced compared to the $e^+e^-$ case. This enhancement is due to the larger Yukawa coupling of the muon.
Moreover, the presence of neutral-Higgs exchange in the
$s$-channel for both processes may enable the possibility of resonance enhancement near $\sqrt{s}\approx m_{h_i,A_j}$. The process $\mu^+ \mu^- \to W^\pm H_1^\mp$ provides the opportunity to search for a charged Higgs with a mass up to $\sqrt{s}-m_W$, unlike
$\mu^+ \mu^- \to H_1^\pm H_1^\mp$, which only allows probing up to
$m_{H^\pm}<\sqrt{s}/2$. 

To do a full study, we have implemented a full set of theoretical constraints originated from \texttt{perturbative unitarity}, \texttt{electroweak vacuum stability}. Also as any BSM scenario must be capable of accommodating withing its spectrum a 125 GeV state $h_{125}$, that exhibits properties consistent with present measurements. To insure that compatibility the \texttt{HiggsTools} package~\cite{Bahl:2022igd} is employed. This guarantees that the allowed parameter space align with the observed properties of the $125$~GeV Higgs boson ( \texttt{HiggsSignals}~\cite{Bechtle:2013xfa,Bechtle:2014ewa,Bechtle:2020uwn,Bahl:2022igd}) and with the limits from searches for additional Higgs bosons at the LHC and at LEP ( \texttt{HiggsBounds}~\cite{Bechtle:2008jh,Bechtle:2011sb,Bechtle:2013wla,Bechtle:2020pkv,Bahl:2022igd}). In addition the parameter space of the model is further constrained using \texttt{Electroweak precision observables} and the \texttt{$\bar{B}\to X_{s}\gamma$ constraint} at 95$\%$ C.L.

\paragraph*{}This paper is organized as follows:  In Sect.\ref{prese_2HDMcT}, we briefly review 2HDMcT model and in Sect.\ref{COMPUTATIONAL PROCEDURE STEPS} we introduce our computational procedure. In Sect.\ref{constraint} we mention the theoretical and experimental constraints imposed on the model parameter space and present our results. Sect.\ref{conlusion} is devoted to our conclusion.

\section{2HDMcT: BRIEF REVIEW}
\label{prese_2HDMcT}
The 2HDMcT contains two Higgs doublets $\Phi_{i}$ (i = 1,2)	and one colorless scalar field $\Delta$ transforming as a triplet under the $SU(2)_L$ gauge group with hypercharge $Y_\Delta=2$. In this case, the most general gauge-invariant Lagrangian of 2HDMcT is given by, 
%
	\begin{equation}
		\begin{matrix}
			\mathcal{L}=\sum_{i=1}^2(D_\mu{\Phi_i})^\dagger(D^\mu{\Phi_i})+Tr(D_\mu{\Delta})^\dagger(D^\mu{\Delta})\vspace*{0.12cm}\\
			\hspace{-3cm}-V(\Phi_i, \Delta)+\mathcal{L}_{\rm Yukawa}
			\label{eq:thdmt-lag}
		\end{matrix}
	\end{equation}
	where the covariant derivatives are defined as,
	\begin{equation}
		D_\mu{\Phi_i}=\partial_\mu{\Phi_i}+igT^a{W}^a_\mu{\Phi_i}+i\frac{g'}{2}B_\mu{\Phi_i} \label{eq:covd1}
	\end{equation}
	\vspace*{-0.6cm}
	\begin{equation}
		D_\mu{\Delta}=\partial_\mu{\Delta}+ig[T^a{W}^a_\mu,\Delta]+ig' \frac{Y_\Delta}{2} B_\mu{\Delta} \label{eq:covd2}
	\end{equation}
	(${W}^a_\mu$, $g$), and ($B_\mu$, $g'$) denoting respectively the $SU(2)_L$ and $U(1)_Y$ gauge fields and couplings and $T^a \equiv \sigma^a/2$, with $\sigma^a$ ($a=1, 2, 3$)  the Pauli matrices. In terms of the two $SU(2)_L$ Higgs doublets $\Phi_i$ and the triplet field $\Delta$, the 2HDMcT scalar potential is given by \cite{Branco:2011iw,Ouazghour:2018mld}:
	
	\begin{widetext}
		\[
		\begin{matrix}
			V(\Phi_1,\Phi_2,\Delta) &=& m_{11}^2 \Phi_1^\dagger\Phi_1+m_{22}^2\Phi_2^\dagger\Phi_2-[m_{12}^2\Phi_1^\dagger\Phi_2+{\rm h.c.}]+\frac{\lambda_1}{2}(\Phi_1^\dagger\Phi_1)^2
			+\frac{\lambda_2}{2}(\Phi_2^\dagger\Phi_2)^2
			\nonumber\\
			&+& \lambda_4(\Phi_1^\dagger\Phi_2)(\Phi_2^\dagger\Phi_1)+ \left\{\frac{\lambda_5}{2}(\Phi_1^\dagger\Phi_2)^2
			+\big[\beta_1(\Phi_1^\dagger\Phi_1)
			+\beta_2(\Phi_2^\dagger\Phi_2)\big]
			\Phi_1^\dagger\Phi_2+{\rm h.c.}\right\} \nonumber\\
			&+&\lambda_3(\Phi_1^\dagger\Phi_1)(\Phi_2^\dagger\Phi_2)+\lambda_6\,\Phi_1^\dagger \Phi_1 Tr\Delta^{\dagger}{\Delta} +\lambda_7\,\Phi_2^\dagger \Phi_2 Tr\Delta^{\dagger}{\Delta}\nonumber\\
			&+&\left\{\mu_1 \Phi_1^T{i}\sigma^2\Delta^{\dagger}\Phi_1 + \mu_2\Phi_2^T{i}\sigma^2\Delta^{\dagger}\Phi_2 + \mu_3 \Phi_1^T{i}\sigma^2\Delta^{\dagger}\Phi_2  + {\rm h.c.}\right\}+\lambda_8\,\Phi_1^\dagger{\Delta}\Delta^{\dagger} \Phi_1\nonumber\\
			&+&\lambda_9\,\Phi_2^\dagger{\Delta}\Delta^{\dagger} \Phi_2+m^2_{\Delta}\, Tr(\Delta^{\dagger}{\Delta}) +\bar{\lambda}_8(Tr\Delta^{\dagger}{\Delta})^2\hspace*{0cm}+\hspace*{0cm}\bar{\lambda}_9Tr(\Delta^{\dagger}{\Delta})^2
			\label{scalar_pot}
		\end{matrix}
		\]
	\end{widetext}
	where $Tr$ denotes the trace over 2x2 matrices. The triplet $\Delta$ and the Higgs doublets $\Phi_{i}$ are represented by, 
	\begin{eqnarray}
		\Delta &=&\left(
		\begin{array}{cc}
			\delta^+/\sqrt{2} & \delta^{++} \\
			(v_t+\delta^0+i\eta_0)/\sqrt{2} & -\delta^+/\sqrt{2}\\
		\end{array}
		\right)\end{eqnarray}
	\vspace*{-0.25cm}
	\begin{eqnarray}
		\Phi_1&=&\left(
		\begin{array}{c}
			\phi_1^+ \\
			\phi^0_1 \\
		\end{array}
		\right){,}~~~\Phi_2=\left(
		\begin{array}{c}
			\phi_2^+ \\
			\phi^0_2 \\
		\end{array}
		\right)\end{eqnarray}
	with $\phi^0_1=(v_1+\rho_1+i\eta_1)/\sqrt{2}$ and $\phi^0_2=(v_2+\rho_2+i\eta_2)/\sqrt{2}$. After the spontaneous electroweak symmetry breaking, the
	Higgs doublets and triplet fields acquire their vacuum expectation values, respectively dubbed $v_1$, $v_2$ and $v_t$, and eleven physical Higgs
	states appear, namely: three CP-even neutral Higgs bosons $(H_1, H_2, H_3)$, four simply charged Higgs bosons $(H_1^{\pm}, H_2^{\pm})$,  two CP odd Higgs $(A_1, A_2)$, and finally two doubly charged Higgs bosons $H^{\pm\pm}$.  
	
	The Yukawa Lagrangian encompasses the entire Yukawa sector of the 2HDM along with an additional Yukawa  term $\mathcal{L}_{\rm Yukawa}$. This term, following spontaneous symmetry breaking, generates (Majorana) mass terms for the neutrinos,
	\begin{equation}
		-\mathcal{L}_{\rm Yukawa} \supset  - Y_{\nu} L^T C \otimes i \sigma^2 \Delta L  + {\rm h.c.} \label{eq:yukawa}
	\end{equation}
	
	\noindent
	\textcolor{black}{
		where $L$ denotes $SU(2)_L$ doublets of left-handed leptons, $Y_{\nu}$ denotes neutrino Yukawa couplings,  
		$C$ the charge conjugation operator. We list in Table-\ref{table1}, all the ${\mathcal{CP}}_{even}$ $h_i$ (i=1,2,3) and ${\mathcal{CP}}_{odd}$ $A_j$ (i=1,2) Yukawa couplings for all Yukawa textures in the model.}

	   \begin{table*}[t]
		\begin{center}
			\begin{tabular}{|c|c|c|c|c|c|c|c|c|c|c|c|c|c|c|c|c|c}
				\hline  & $C^{h_1}_U$    & $C^{h_1}_D$ & $C^{h_1}_l$   &   $C^{h_2}_U$   &   $C^{h_2}_D$ &   $C^{h_2}_l$   &   $C^{h_3}_U$  &   $C^{h_3}_D$ &   $C^{h_3}_l$  & $C^{A_1}_U$  & $C^{A_1}_D$ & $C^{A_1}_l$ & $C^{A_2}_U$  & $C^{A_2}_D$ & $C^{A_2}_l$ \\
				\hline  Type-I & $\displaystyle\frac{\mathcal{E}_{12}}{s_\beta} $ & $\displaystyle\frac{\mathcal{E}_{12}}{s_\beta} $&
				$\displaystyle\frac{\mathcal{E}_{12}}{s_\beta} $& $\displaystyle\frac{\mathcal{E}_{22}}{s_\beta} $ & $\displaystyle\frac{\mathcal{E}_{22}}{s_\beta} $ &
				$\displaystyle\frac{\mathcal{E}_{22}}{s_\beta} $ & $\displaystyle\frac{\mathcal{E}_{32}}{s_\beta} $ & $\displaystyle\frac{\mathcal{E}_{32}}{s_\beta} $ &
				$\displaystyle\frac{\mathcal{E}_{32}}{s_\beta} $ &
				
				$\displaystyle\frac{\mathcal{O}_{22}}{s_\beta} $  &
				$\displaystyle\frac{\mathcal{O}_{22}}{s_\beta} $ &
				$\displaystyle\frac{\mathcal{O}_{22}}{s_\beta} $ & $\displaystyle\frac{\mathcal{O}_{32}}{s_\beta} $& $\displaystyle\frac{\mathcal{O}_{32}}{s_\beta} $& $\displaystyle\frac{\mathcal{O}_{32}}{s_\beta} $
				\\
				\hline  Type-II &$\displaystyle\frac{\mathcal{E}_{12}}{s_\beta} $& $\displaystyle\frac{\mathcal{E}_{11}}{c_\beta} $ &
				$\displaystyle\frac{\mathcal{E}_{11}}{c_\beta} $ & $\displaystyle\frac{\mathcal{E}_{22}}{s_\beta} $ & $\displaystyle\frac{\mathcal{E}_{21}}{c_\beta} $ &
				$\displaystyle\frac{\mathcal{E}_{21}}{c_\beta} $ & $\displaystyle\frac{\mathcal{E}_{32}}{s_\beta} $ & $\displaystyle\frac{\mathcal{E}_{31}}{c_\beta} $ 
				&$\displaystyle\frac{\mathcal{E}_{31}}{c_\beta} $
				
				&$\displaystyle\frac{\mathcal{O}_{22}}{s_\beta} $&$\displaystyle\frac{\mathcal{O}_{21}}{c_\beta} $
				&$\displaystyle\frac{\mathcal{O}_{21}}{c_\beta} $&$\displaystyle\frac{\mathcal{O}_{32}}{s_\beta} $&$\displaystyle\frac{\mathcal{O}_{31}}{c_\beta} $&$\displaystyle\frac{\mathcal{O}_{31}}{c_\beta} $
				\\
				\hline  Type-III &$\displaystyle\frac{\mathcal{E}_{12}}{s_\beta} $& $\displaystyle\frac{\mathcal{E}_{12}}{s_\beta} $ &
				$\displaystyle\frac{\mathcal{E}_{11}}{c_\beta} $ & $\displaystyle\frac{\mathcal{E}_{22}}{s_\beta} $ & $\displaystyle\frac{\mathcal{E}_{22}}{s_\beta} $ &
				$\displaystyle\frac{\mathcal{E}_{21}}{c_\beta} $ & $\displaystyle\frac{\mathcal{E}_{32}}{s_\beta} $ &$\displaystyle\frac{\mathcal{E}_{32}}{s_\beta} $
				&$\displaystyle\frac{\mathcal{E}_{31}}{c_\beta} $&
				$\displaystyle\frac{\mathcal{O}_{22}}{s_\beta} $&
				$\displaystyle\frac{\mathcal{O}_{22}}{s_\beta} $&
				$\displaystyle\frac{\mathcal{O}_{21}}{c_\beta} $ &
				$\displaystyle\frac{\mathcal{O}_{32}}{s_\beta} $& $\displaystyle\frac{\mathcal{O}_{32}}{s_\beta} $& 
				$\displaystyle\frac{\mathcal{O}_{31}}{c_\beta} $ 
				
				\\ 
				\hline  Type-IV & $\displaystyle\frac{\mathcal{E}_{12}}{s_\beta} $ & $\displaystyle\frac{\mathcal{E}_{11}}{c_\beta} $&
				$\displaystyle\frac{\mathcal{E}_{12}}{s_\beta} $& $\displaystyle\frac{\mathcal{E}_{22}}{s_\beta} $ & $\displaystyle\frac{\mathcal{E}_{21}}{c_\beta} $ &
				$\displaystyle\frac{\mathcal{E}_{22}}{s_\beta} $ & $\displaystyle\frac{\mathcal{E}_{32}}{s_\beta} $ & $\displaystyle\frac{\mathcal{E}_{31}}{c_\beta} $ &
				$\displaystyle\frac{\mathcal{E}_{32}}{s_\beta} $&
				$\displaystyle\frac{\mathcal{O}_{22}}{s_\beta} $ & $\displaystyle\frac{\mathcal{O}_{21}}{c_\beta} $ &
				$\displaystyle\frac{\mathcal{O}_{22}}{s_\beta} $& $\displaystyle\frac{\mathcal{O}_{32}}{s_\beta} $ &$\displaystyle\frac{\mathcal{O}_{31}}{c_\beta} $&$\displaystyle\frac{\mathcal{O}_{32}}{s_\beta} $
				\\  
				\hline
			\end{tabular}
		\end{center} 
		\caption{Normalized Yukawa couplings coefficients of the neutral Higgs bosons $h_i$ and $A_j$ to the leptons, up and down quarks ($u,d$) in 2HDMcT. For the expressions of the coefficients $\displaystyle{\mathcal{E}_{ij}}$ and  $\displaystyle{\mathcal{O}_{ij}}$ see Ref~\cite{Ouazghour:2018mld}.}
		\label{table1}
	\end{table*}

	On the other hand, expanding the covariant derivative ${\rm D}_\mu$, and performing the usual transformations on the gauge and scalar fields to obtain the physical fields, one can identify the Higgs couplings of $h_i$\footnote{Since we are working in the CP-Conserving scenario the couplings of the CP-odd Higgs bosons to the gauge bosons are zero.} to the massive gauge bosons $V=W,Z$ as given in Table-\ref{table2}. Note that in our model, the triplet field $\Delta$ does directly couple to the SM particles, so a new contribution will appear, and the two couplings $C^{h_i}_V$ ($V=W^\pm, Z$) differs from one to another by a factor 2 associated to $v_t$.
	
	\begin{table}[!h]
		\begin{center}
			\begin{tabular}{|c|c|c|}
				\hline  
				& $C^{h_i}_W$    & $C^{h_i}_Z$  \\
				\hline  $h_1$ & $\displaystyle{\frac{v_1}{v} \mathcal{E}_{11} + \frac{v_2}{v} \mathcal{E}_{21} + 2\,\frac{v_t}{v} \mathcal{E}_{31}}$ & 
				$\displaystyle{\frac{v_1}{v} \mathcal{E}_{11} + \frac{v_2}{v} \mathcal{E}_{21} + 4\,\frac{v_t}{v} \mathcal{E}_{31}}$  \\
				\hline  $h_2$ & $\displaystyle{\frac{v_1}{v} \mathcal{E}_{12} + \frac{v_2}{v} \mathcal{E}_{22} + 2\,\frac{v_t}{v} \mathcal{E}_{32}}$ & 
				$\displaystyle{\frac{v_1}{v} \mathcal{E}_{12} + \frac{v_2}{v} \mathcal{E}_{22} + 4\,\frac{v_t}{v} \mathcal{E}_{32}}$  \\
				\hline  $h_3$ & $\displaystyle{\frac{v_1}{v} \mathcal{E}_{13} + \frac{v_2}{v} \mathcal{E}_{23} + 2\,\frac{v_t}{v} \mathcal{E}_{33}}$ & 
				$\displaystyle{\frac{v_1}{v} \mathcal{E}_{13} + \frac{v_2}{v} \mathcal{E}_{23} + 4\,\frac{v_t}{v} \mathcal{E}_{33}}$  \\
				\hline 
			\end{tabular}
			\caption{The normalized couplings of the neutral ${\mathcal{CP}}_{even}$ $h_i$ Higgs bosons
				to the massive gauge bosons $V=W,Z$ in 2HDMcT.}
			\label{table2}
		\end{center}
	\end{table}
	\section{COMPUTATIONAL PROCEDURE STEPS}
	\label{COMPUTATIONAL PROCEDURE STEPS}
	In this section, we list all the 2-to-2 processes under investigation in this study. We will provide the contributing tree-level Feynman diagrams, along with their corresponding amplitudes and squared amplitudes.
	\subsubsection{{\color{blue}Charged Higgs pair production and $\mu^+ \mu^- \to H_1^{\pm}H_2^{\mp}$ processes}}
	\label{section3subsec1}
	%
	\begin{figure*}[htbp]
		\centering
		\includegraphics[width=0.45\textwidth]{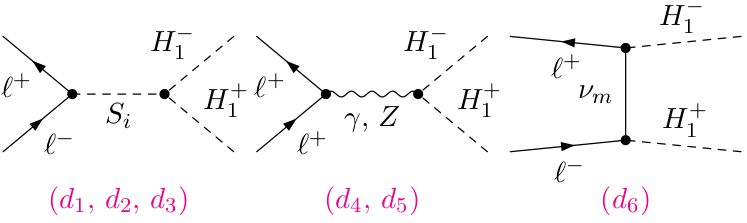}%
		\caption{Tree-level Feynman diagrams for $\mu^+ \mu^- \to H_1^{+}H_1^{-}$ at muon collider in the 2HDMcT. $S_i=h_1,h_2,h_3$.}
		\label{diag:2mu2Hp1Hm1}
	\end{figure*}
	\begin{figure*}[htbp]
		\centering
		\includegraphics[width=0.45\textwidth]{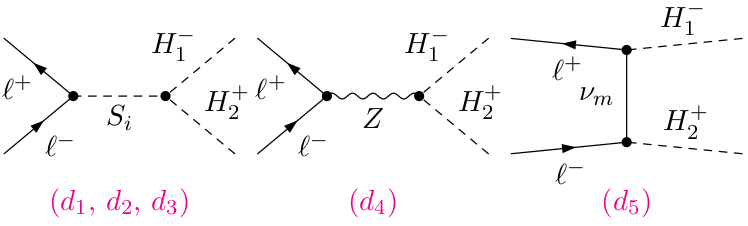}%
		\caption{Tree-level Feynman diagrams for $\mu^+ \mu^- \to H_1^{+}H_2^{-}$ at muon collider in the 2HDMcT. $S_i=h_1,h_2,h_3$.}
		\label{diag:2mu2Hp1Hm2}
	\end{figure*}
	\noindent
	The tree-level Feynman diagrams contributing to $\mu^{+} \mu^{-} \rightarrow H_1^{+}H_1^{-} $ in the 2HDMcT are given in Fig.~\ref{diag:2mu2Hp1Hm1}. At tree level the pair production process receives contributions from the conventional Drell--Yan mechanism $\mu^+\mu^-\to \gamma^*,Z^* \to H_1^+H_1^-$ in Fig. \ref{diag:2mu2Hp1Hm1}$(d_{4,5})$, as well as the $s$-channel neutral-Higgs exchange $\mu^+\mu^-\to h_1^*, h_2^*, h_3^* \to H_1^+H_1^-$ in Fig. \ref{diag:2mu2Hp1Hm1}$(d_{1,2,3})$, and the $t$-channel neutrino-exchange diagram in Fig. \ref{diag:2mu2Hp1Hm1}$(d_{6})$. The $s$-channel neutral-Higgs exchange diagrams can largely be enhanced near the resonance region $\sqrt s \approx m_{h_i}$. The $t$-channel diagram in Fig. \ref{diag:2mu2Hp1Hm1} experiences suppression due to the Yukawa coupling at two vertices. Nevertheless, if the Yukawa coupling of the charged Higgs to the muon becomes significant, it could result in an enhancement at high center-of-mass energies.
	
	The process $\mu^+\mu^- \to H_1^\pm H_1^\mp$ is indicated by,
	\begin{equation} \label{mumutohphm}
		\mu^+(p_1) \mu^-(p_2) \rightarrow H_1^\pm(k_1) H_1^\mp(k_2),
	\end{equation}
	Where the 4-momenta of the incoming $\mu^+$ and $\mu^-$, as well as the outgoing charged Higgs boson $H_1^{\pm}$ and $H_1^{\mp}$, are denoted in parentheses. The  four-momenta are defined in the center-of-mass frame as :
	\begin{eqnarray}
		p_{1,2} & = & \big(\frac{\sqrt{s}}{2},\,0\,,0\,, \pm \frac{\sqrt{s}}{2} \big) \label{2mu2HpHm:p1}\nonumber\\
		k_{1,2} & = & \big(\frac{\sqrt{s}}{2},\,\pm \frac{\sqrt{s}}{2}\,\beta\sin\theta,0,\,\pm \frac{\sqrt{s}}{2}\,\beta\cos\theta\big)\;, \label{2mu2HpHm:k1}
	\end{eqnarray}
	where $\beta = \sqrt{1-4\,m_{H_1^\pm}^2/s}$, $\sqrt{s}$ represents the center-of-mass energy, and $\theta$ denotes the scattering angle between $\mu^+$ and $H_1^+$ in the center-of-mass frame.
	
	\noindent
	The Mandelstam variables $s, t $ and $u$  are defined as follows :
	\begin{eqnarray}
		s & = & \big(p_{1}+p_{2}\big)^2 = \big(k_{1}+k_{2}\big)^2 \label{2mu2HpHm:s}\nonumber\\
		t & = & \big(p_{1}-k_{1}\big)^2 = \big(p_{2}-k_{2}\big)^2 = m_{H_1^\pm}^2 -\frac{s}{2} +\frac{s}{2} \beta \cos\theta \label{2mu2HpHm:t}\nonumber\\
		u & = & \big(p_{1}-k_{2}\big)^2 = \big(p_{2}-k_{1}\big)^2 = m_{H_1^\pm}^2 -\frac{s}{2} -\frac{s}{2} \beta \cos\theta  \label{2mu2HpHm:u} \;.
	\end{eqnarray}
	
	Using the Feynman rules, the matrix elements for this process are as follows :

	\begin{eqnarray}
		M_{Tree}^{\gamma} &=&    -2\frac{ e^2(\mathcal{C}_{21}^2+\mathcal{C}_{22}^2+\mathcal{C}_{23}^2)}{s}   \bar{v}(p_2) \slashed{k}_2 u(p_1) \nonumber \\
		M_{Tree}^{z} &=&  \frac{2\lambda_{ZH_1^{\pm}H_1^{\mp}} }{s-m_Z^2+i m_Z \Gamma_Z} \bar{v}(p_2)  \slashed{k}_2 \big(g_V-g_A\gamma^5\big) u(p_1)   \nonumber\\
		M_{Tree}^{\nu} &=& - \frac{g^2 m_{\mu}^2 Y_6^2 }{4 m_W^2  t  } [ \bar{v}(p_2)  \slashed{k}_2  (1+\gamma_5) u(p_1)   +   
		m_{\mu} \bar{v}(p_2) (1+\gamma_5)   u(p_1)  ]  \nonumber  \\
		M_{Tree}^{h_i} &=&\bar{v}(p_2) u(p_1)\frac{\lambda_{h_iH_1^+H_1^-}}{s-m_{h_i}^2+i m_{h_i} \Gamma_{h_i}} \frac{g m_{\mu} C_\ell^{h_i}}{2 m_W} 
	\end{eqnarray}
	where the following couplings have been used :
	$\lambda_{Z^\mu\mu^+\mu^-} = i\,\gamma^\mu\big(g_V-g_A\gamma^5\big)$ 
	$g_V= g (1-4\,s_w^2)/(4\,c_w)$ and $g_A=g/(4\,c_w)$.
	
	\noindent
	The $Z$ coupling to the pair of charged Higgs $H_1^{\pm}$ is : 
	$\lambda_{ZH_1^{\pm}H_1^{\mp}} =-g  [(c_w^2-s_w^2)(C_{21}^2+C_{22}^2)-2 s_w^2 C_{23}^2]/2c_w$.\\
	
	The total widths for the $Z$ boson, $h_1$, $h_2$ and $h_3$ are introduced because of their necessity for the case where the center-of-mass energy becomes close to the mass of the neutral-Higgs state : $\sqrt{s}\approx m_{h_i}$. The total widths for $h_1$, $h_2$ and $h_3$ are computed at the leading order. We neglect the muon-mass term in $M_{Tree}^{\nu}$ amplitude.
	
	\noindent
	The amplitude squared is then given by :
	\begin{align}
		|M|^2 &=& \Big[ \big(|a_V|^2 + |a_A|^2 \big) \frac{s^2}{2}\beta^2 \sin^2\theta  + 2\,\big(|a_S|^2 -|a_{SA}|^2\big)\,s \Big] \;,
		\label{Amsquare}
	\end{align}
	Where,
	
	\begin{eqnarray}
		a_V & = &   \frac{2 \lambda_{ZH_1^{\pm}H_1^{\mp}} g_V}{s-M_Z^2+ i M_Z \Gamma_Z} - 2(\mathcal{C}_{21}^2+\mathcal{C}_{22}^2+\mathcal{C}_{23}^2) \frac{e^2}{s}  \nonumber  \\
		 &-& \frac{g^2 m_\mu^2  Y_6^2 }{4 M_W^2t} \nonumber  \\
		a_A & = &- \frac{2  \lambda_{ZH_1^{\pm}H_1^{\mp}} g_A}{s-M_Z^2+ i M_Z \Gamma_Z}  - \frac{g^2 m_\mu^2  Y_6^2 }{4 M_W^2t}  \nonumber \\
		a_S & = &  \frac{ g m_\mu }{2  M_W }\sum_{i=1}^{3}\frac{\lambda_{h_i H_1^+ H_1^-} C_\ell^{h_i}}{s-M_{h_i}^2+i M_{h_i} \Gamma_{h_i}} - \frac{g^2 m_\mu^3  Y_6^2 }{4 M_W^2t} \nonumber \\
		a_{SA} & = & - \frac{g^2 m_\mu^3  Y_6^2 }{4 M_W^2t} 
	\end{eqnarray}
	
	It is evident from Eq. \ref{Amsquare} that the squared amplitude of the $s$-channel neutral-Higgs ($h_i,\ i=1,2,3$) exchange is independent of the scattering angle, resulting in a flat angular distribution for these contributions.
	\noindent
	The differential cross section is given by:
	\begin{eqnarray}
		\frac{d\sigma}{d\Omega}=\frac{1}{4}\frac{\beta}{64 \pi^2 s} \;.
		|M|^2 \label{ref0}
	\end{eqnarray}
	$1/4$ is a factor due to initial state spin average. 
	
	In the case of $e^+e^-$ collision, contributions from the s-channel with $h_1$, $h_2$ and $h_3$ exchange and the t-channel neutrino exchange are neglected due to their proportionality to the electron mass. The total cross section for $e^+e^- \to H_1^+ H_1^-$ 
	is as follows \cite{Arhrib:1998gr}:
	\begin{equation}
		\label{eq:tot_cs_ee_hphm}
		\sigma_{\rm tot}^{e^+e^-}=\frac{e^4 \pi\alpha^2\beta^3}{3s}\bigg( 1 + { \lambda_{ZH_1^{\pm}H_1^{\mp}}}^2 \frac{g_V^2+g_A^2}{(1-m_Z^2/s)^2}-\frac{2 \lambda_{ZH_1^{\pm}H_1^{\mp}}g_V}{1-m_Z^2/s} \bigg)
	\end{equation}
	\vspace{6pt}
	Similarly to the previous process, the process $\mu^+\mu^- \to H_1^\pm H_2^{\mp}$\footnote{The process $\mu^+\mu^- \to H_1^\pm H_2^{\mp}$ is absent in 2HDM.} can be indicated by,
	\begin{equation} \label{eq:gammah}
		\mu^+(p_1) \mu^-(p_2) \rightarrow H_1^\pm(k_1) H_2^{\mp}(k_2),
	\end{equation}
	Where the 4-momenta of the incoming $\mu^+$ and $\mu^-$, as well as the outgoing charged Higgs boson $H_1^{\pm}$ and  $H_2^{\mp}$, are denoted in parentheses. 
	
	\noindent
	The Mandelstam variables can be written as :
	\begin{equation}
		s=(p_1+p_2)^2, \quad t=(p_1-k_1)^2,\quad
		u=(p_1-k_2)^2.
	\end{equation}
	Neglecting the muon mass $m_{\mu}$, the momenta in the $\mu^+ \mu^-$ center-of-mass system are expressed as follows :
	\begin{eqnarray}
		& &p_{1,2} = \frac{\sqrt{s}}{2} (1,0,0,\pm 1) \nonumber \\
		& &k_{1, 2} = \frac{\sqrt{s}}{2} \left(1 \pm \frac{m_{H_2^{\mp}}^2 - m_{H_1^\pm}^2}{s}, \pm \frac{1}{s} \lambda^{\frac{1}{2}}(s,m_{H_2^{\mp}}^2,m_{H_1^\pm}^2) \sin\theta,0, \right. \nonumber \\
		& &
		\left. \pm \frac{1}{s} \lambda^{\frac{1}{2}}(s,m_{H_2^{\mp}}^2,m_{H_1^\pm}^2) \cos\theta \right), \nonumber
	\end{eqnarray}
	\noindent
	Where $\lambda(x,y,z)$ is the usual two body phase space function given by,
	\begin{eqnarray}
		\lambda(x,y,z)=x^2+y^2+z^2-2xy-2xz-2yz
	\end{eqnarray}
	And $\theta$ is the scattering angle between $\mu^+$ and $H_1^+$. The $s,\, t$ and $u$ can be written as :
	\begin{eqnarray}
		s &=&  (p_{1}+p_{2})^2 = (k_1+k_{2})^2  \nonumber\\
		t &=& (p_{1}-k_1)^2 = (p_{2}-k_{2})^2 =
		\frac{1}{2}(m_{H_2^{\mp}}^2  + m_{H_1^\pm}^2) -
		\frac{s}{2} \nonumber \\
		& &
		+\frac{1}{2} \lambda^{\frac{1}{2}}(s,m_{H_2^{\mp}}^2,m_{H_1^\pm}^2)
		\cos\theta  \nonumber\\
		u &=& (p_{1}-k_{2})^2 = (p_{2}-k_1)^2 =
		\frac{1}{2} (m_{H_2^{\mp}}^2 + m_{H_1^\pm}^2) -
		\frac{s}{2}\nonumber \\
		& &
		-\frac{1}{2} \lambda^{\frac{1}{2}}(s,m_{H_2^{\mp}}^2,m_{H_1^\pm}^2) \cos\theta \\
		s&+&t+u = m_{H_2^{\mp}}^2 + m_{H_1^\pm}^2  \nonumber
	\end{eqnarray}
	The $s$-channel $M_{Tree}^{h_i}, i=1,2,3$, $M_{Tree}^{A_j} j=1,2$  
	and $t$-channel $M_{Tree}^{\nu}$ amplitudes are respectively given by :
	\begin{eqnarray}
		M_{Tree}^{z} &=&  \frac{2\lambda_{ZH_1^{\pm}H_2^{\mp}} }{s-m_Z^2+i m_Z \Gamma_Z} \bar{v}(p_2)  \slashed{k}_2 \big(g_V-g_A\gamma^5\big) u(p_1)   \nonumber\\
		M_{Tree}^{\nu} &=& - \frac{g^2 m_{\mu}^2 Y_6Y_7 }{2 m_W^2  t  } [ \bar{v}(p_2)  \slashed{k}_2 (1+\gamma_5) u(p_1) \nonumber\\
		&+& m_{\mu} \bar{v}(p_2) (1+\gamma_5) u(p_1)]  \nonumber  \\
		M_{Tree}^{h_i} &=&\bar{v}(p_2) u(p_1)\frac{\lambda_{h_iH_1^+H_2^-}}{s-m_{h_i}^2+i m_{h_i} \Gamma_{h_i}} \frac{g m_{\mu} C_\ell^{h_i}}{2 m_W} \nonumber  \\
	\end{eqnarray}
	\begin{eqnarray*}
		|M|^2 &=&\frac{1}{2} \Big[-\lambda(s,m_{H_2^\pm}^2,m_{H_1^\pm}^2) \big(|a_1|^2 + |a_2|^2 \big) \cos^2\theta  \nonumber  \\
		&+& \big(|a_1|^2 + |a_2|^2 \big) \big(\big( m_{H_1^\pm}-m_{H_2^\pm}\big)^2-s\big)\Big. \nonumber \\
		&&\Big.\big(\big( m_{H_1^\pm}+m_{H_2^\pm}\big)^2-s\big)+4s\big(|a_3|^2+|a_4|^2\big)\Big]\;\;\;\;\;\;\;\;\;\;\;\;\;\;\;\;\;\;\;\;\;\;\;\;\;\;\;\;\;\;\;\;\;\;\;\;\;\;\;\;\;\;\;\;\;\;\;\;\;\;\;\;\;\;\;\;\; \;
		\label{Amsquare_}
	\end{eqnarray*}
	\begin{eqnarray}
		a_1 & = &   \frac{2 \lambda_{ZH_1^{\pm}H_2^{\mp}} g_V}{s-M_Z^2+ i M_Z \Gamma_Z}  - \frac{g^2 m_\mu^2 Y_6Y_7 }{4 M_W^2t} \nonumber  \\
		a_2 & = &- \frac{2  \lambda_{ZH_1^{\pm}H_2^{\mp}} g_A}{s-M_Z^2+ i M_Z \Gamma_Z}  - \frac{g^2 m_\mu^2 Y_6Y_7 }{4 M_W^2t}  \nonumber \\
		a_3 & = &  \frac{ g m_\mu }{2  M_W }\sum_{i=1}^{3}\frac{\lambda_{h_i H_1^+ H_2^-} C_\ell^{h_i}}{s-M_{h_i}^2+i M_{h_i} \Gamma_{h_i}} - \frac{g^2 m_\mu^3 Y_6Y_7 }{4 M_W^2t} \nonumber \\
		a_{4} & = & - \frac{g^2 m_\mu^3 Y_6Y_7 }{4 M_W^2t} 
	\end{eqnarray}
	\subsubsection{\textcolor{blue}{$\mu^+ \mu^- \to H_1^{\pm}W^{\mp}$ process}}
	\label{section3}
\begin{figure*}[!htb]
		\centering
		\includegraphics[width=0.65\textwidth]{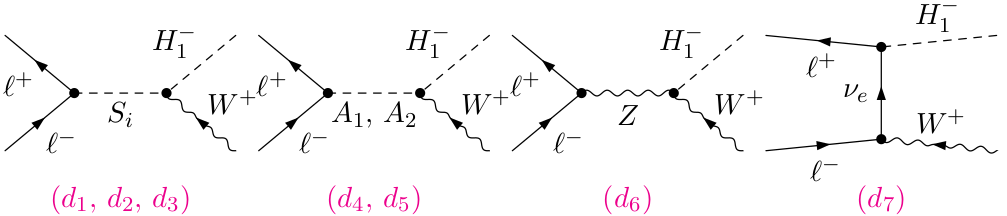}%
		\caption{Tree-level Feynman diagrams for $\mu^+ \mu^- \to H_1^{\pm}W^{\mp}$ at a muon collider in the 2HDMcT. $S_i=h_1,h_2,h_3$.}
		\label{diag:2mu2HpWm}
	\end{figure*}

	In the 2HDMcT model, the process $\mu^+ \mu^- \to H_1^{\pm}W^{\mp}$ can occur through an $s$-channel mediated by the neutral Higgs bosons $h_1, h_2, h_3$, $A_1, A_2$ and $Z$ (see Fig. \ref{diag:2mu2HpWm} - $(d_{1,2,3,4,5,6})$) and through a $t$-channel diagram involving neutrino exchange (see Fig. \ref{diag:2mu2HpWm} - $(d_7)$). This process at a muon collider is highly promising for several reasons. Firstly, it can provide insights into the underlying Yukawa textures, as all the diagrams shown in Fig. \ref{diag:2mu2HpWm} involve the muon coupling. Secondly, the process may experience resonance enhancement through $s$-channel heavy-Higgs exchanges of $h_{1,2,3}$ and $A_{1,2}$ via the following pathways: $\mu^+ \mu^- \to h_{1,2,3}^*, A_{1,2}^* \to H_1^{\pm}W^{\mp}$.

	The process $\mu^+\mu^- \to H_1^\pm W^\mp$ is indicated by,
	\begin{equation} \label{eq:gammah}
		\mu^+(p_1) \mu^-(p_2) \rightarrow H_1^\pm(k_1) W^\mp(k_2),
	\end{equation}
	Where the 4-momenta of the incoming $\mu^+$ and $\mu^-$, as well as the outgoing charged Higgs boson $H_1^{\pm}$ and the gauge boson $W^{\mp}$, are denoted in parentheses. The Mandelstam variables can be written as :
	\begin{equation}
		s=(p_1+p_2)^2, \quad t=(p_1-k_1)^2,\quad
		u=(p_1-k_2)^2.
	\end{equation}
	Neglecting the muon mass $m_{\mu}$, the momenta in the $\mu^+ \mu^-$ center-of-mass system are expressed as follows :
	\begin{eqnarray}
		& &p_{1,2} = \frac{\sqrt{s}}{2} (1,0,0,\pm 1) \nonumber \\
		& &k_{1, 2} = \frac{\sqrt{s}}{2} \left(1 \pm \frac{m_W^2 - m_{H_1^\pm}^2}{s}, 
		\pm \frac{1}{s} \lambda^{\frac{1}{2}}(s,m_W^2,m_{H_1^\pm}^2) \sin\theta,0,  \right. \nonumber \\
		& &
		\left.\pm \frac{1}{s} \lambda^{\frac{1}{2}}(s,m_W^2,m_{H_1^\pm}^2) \cos\theta \right), \nonumber
	\end{eqnarray}
	\noindent
	Where $\lambda(x,y,z)$ is the usual two body phase space function given by,
	\begin{eqnarray}
		\lambda(x,y,z)=x^2+y^2+z^2-2xy-2xz-2yz
	\end{eqnarray}
	And $\theta$ is the scattering angle between $\mu^+$ and $H_1^+$. The $s,\, t$ and $u$ can be written as :
	\begin{eqnarray}
		s &=&  (p_{1}+p_{2})^2 = (k_1+k_{2})^2  \nonumber\\
		t &=& (p_{1}-k_1)^2 = (p_{2}-k_{2})^2 =
		\frac{1}{2}(m_W^2  + m_{H_1^\pm}^2) -\frac{s}{2} \nonumber\\
		&+&
		\frac{1}{2} \lambda^{\frac{1}{2}}(s,m_W^2,m_{H_1^\pm}^2)
		\cos\theta  \nonumber\\
		u &=& (p_{1}-k_{2})^2 = (p_{2}-k_1)^2 =
		\frac{1}{2} (m_W^2 + m_{H_1^\pm}^2) -
		\frac{s}{2}\nonumber\\
		&-&
		\frac{1}{2} \lambda^{\frac{1}{2}}(s,m_W^2,m_{H_1^\pm}^2) \cos\theta \\
		s&+&t+u = m_W^2 + m_{H_1^\pm}^2  \nonumber
	\end{eqnarray}
	The $s$-channel $M_{Tree}^{h_i}, i=1,2,3$, $M_{Tree}^{A_j} j=1,2$  
	and $t$-channel $M_{Tree}^{\nu}$ amplitudes are respectively given by :
	\begin{widetext}
	\begin{eqnarray}
		M_{Tree}^{h_i} &=&   \frac{ g^2 m_{\mu} }{4 m_W  } \frac{\lambda_{h_iH_1^{\pm}W^{\mp}}C_{h_i}^l }{s-m_{h_i}^2 + i m_{h_i} \Gamma_{h_i}}  
		\bar{v}(p_2)  u(p_1) (2 k_1 +k_2)^\mu \epsilon_{\mu}(k_2) \nonumber  \\
		M_{Tree}^{A_j} &=&    \frac{ g^2 m_{\mu} }{4 m_W  } \frac{\lambda_{A_jH_1^{\pm}W^{\mp}} C_{A_j}^l}{s-m_{A_j}^2 + i m_{A_j} \Gamma_{A_j}}  
		\bar{v}(p_2)\gamma_5 u(p_1) (2 k_1 +k_2)^\mu \epsilon_{\mu}(k_2)
		\nonumber  \\
		M_{Tree}^{z} &=&  \frac{\lambda_{ZH_1^{\pm}W^{\mp}} }{s-m_Z^2+i m_Z \Gamma_Z} \bar{v}(p_2) \gamma^\mu  \big(g_V-g_A\gamma^5\big) u(p_1)\epsilon_{\mu}(k_2)  
		\nonumber  \\
		M_{Tree}^{\nu} &=&    \frac{ g^2 m_{\mu} }{4 m_W  } \frac{Y_6}{t}  
		\bar{v}(p_2) \gamma^\mu (1-\gamma^5)  ( \slashed{k}_2- \slashed{p}_2 ) u(p_1)  \epsilon_{\mu}(k_2)  
	\end{eqnarray}

	Taking into account the spin average of the initial state and polarization sum of the $W$ gauge boson,  
	The square of the amplitude is given by~\cite{Mertig:1990an,Shtabovenko:2016sxi,Shtabovenko:2020gxv,Shtabovenko:2023idz} :
	\begin{align}\small
		&|{\cal M}|^2 = \frac{1}{2 M_W^2} \Bigg[ 
		S^2 \Big( -4 m_{H_1^\pm}^2 \big(2(b_1^2 + b_2^2) - 2 b_5 (b_1 + b_2) + b_5^2\big) - 8 M_W^2 \big(b_1^2 - b_5(b_1 + b_2) + b_2^2\big) + b_3^2 + b_4^2 \Big) \nonumber \\
		&\quad - 2 S \Big( M_W^2 \Big(2 m_{H_1^\pm}^2 \big(2(b_1^2 + b_2^2) - 2 b_5 (b_1 + b_2) + b_5^2\big) - 3 (b_3^2 + b_4^2)\Big) + m_{H_1^\pm}^2 \Big( - m_{H_1^\pm}^2 \big(2(b_1^2 + b_2^2) - 2 b_5(b_1 + b_2) + b_5^2\big) + b_3^2 + b_4^2 \Big) \nonumber \\
		&\quad + M_W^4 \Big( -2(b_1^2 + b_2^2) + 2 b_5 (b_1 + b_2) + 3 b_5^2 \Big) \Big) + 2 S^3 \Big( 2(b_1^2 + b_2^2) - 2 b_5 (b_1 + b_2) + b_5^2 \Big) \nonumber \\
		&\quad + \lambda(s, m_{H_1^\pm}^2, M_W^2) \cos\theta \Big( -4 b_5 S (b_1 + b_2 - b_5) (m_{H_1^\pm}^2 + M_W^2 - S) - \lambda(s, m_{H_1^\pm}^2, M_W^2) \cos\theta \big( b_3^2 + b_4^2 + 4 b_5^2 M_W^2 - 2 b_5^2 S \big) \Big) \nonumber \\
		&\quad + (m_{H_1^\pm}^2 - M_W^2)^2 \big( b_3^2 + b_4^2 + 4 b_5^2 M_W^2 \big)
		\Bigg]
	\end{align}
	\begin{eqnarray}
		b_1 & = &   \frac{ g^2 m_{\mu} }{4 m_W  } \sum_{i=1}^{3}\frac{\lambda_{h_iH_1^{\pm}W^{\mp}}C_{h_i}^l }{s-m_{h_i}^2 + i m_{h_i} \Gamma_{h_i}} ,\
		b_2  = \frac{ g^2 m_{\mu} }{4 m_W  } \sum_{j=1}^{2}\frac{\lambda_{A_jH_1^{\pm}W^{\mp}}C_{A_j}^l }{s-m_{A_j}^2 + i m_{A_j} \Gamma_{A_j}}  \nonumber \\
		b_3 & = &  \frac{g_V\lambda_{ZH_1^{\pm}W^{\mp}} }{s-m_Z^2+i m_Z \Gamma_Z},\
		b_{4} =-  \frac{g_A\lambda_{ZH_1^{\pm}W^{\mp}} }{s-m_Z^2+i m_Z \Gamma_Z},\
		b_{5} = \frac{ g^2 m_{\mu} }{4 m_W  } \frac{Y_6}{t}  
	\end{eqnarray}
    \end{widetext}
	The differential cross-section for
	$\sigma(\mu^+\mu^-\to H_1^{\pm}W^{\mp})$ may be written as follows :
	\begin{equation}
		{d\sigma\over d\Omega} = {\lambda^{1\over 2}(s,m_{H^{\pm}}^2,m_W^2)\over
			64\pi^2s^2} |{\cal M}|^2 \label{ref1}
	\end{equation}
	\vspace{6pt}
	\section{Constraints and numerical results}
	\vspace{6pt}
	\label{constraint}
	\paragraph*{}
	The phenomenological analysis in 2HDMcT is performed via  implementation of a full set of theoretical constraints ~\cite{Ouazghour:2018mld, Ouazghour:2023eqr} as well as the Higgs exclusion limits from various experimental measurements at colliders, namely :
	\begin{itemize}
		\item \textbf{Unitarity} : The scattering processes must obey  perturbative unitarity.
		\item \textbf{Perturbativity}: The quartic couplings of the scalar potential are constrained by the following conditions : $| \lambda_i|<8 \pi$ for each $i=1,..,5$.
		\item \textbf{Vacuum stability} : Boundedness from below $BFB$ arising from the positivity in any direction of the fields $\Phi_i$, $\Delta$.
		\item[\textbullet]{\bf Electroweak precision observables}: The oblique parameters $S, T$ and $U$~\cite{Peskin:1991sw,Grimus:2008nb} have been calculated in 2HDMcT~\cite{Ouazghour:2023eqr}.  The analysis of the precision electroweak data in light of the new PDG mass of the $W$ boson yields~\cite{ParticleDataGroup:2024cfk}:	
		\begin{eqnarray}
			\widehat S_0= -0.05\pm 0.07,\  \ \widehat T_0 = 0.00\pm 0.06,\ \ \rho_{ST} = 0.93,  \nonumber 
		\end{eqnarray}	
		We use the following $\chi^2_{ST}$ test :
		
		\begin{equation}
			\small
			\label{eq:STRange}
			\frac{(S-\widehat S_0)^2}{\sigma_S^2}\ +\
			\frac{(T-\widehat T_0)^2}{\sigma_T^2}\ -\
			2\rho_{ST}\frac{(S-\widehat S_0)(T-\widehat T_0)}{\sigma_S \sigma_T}\
			\leq\ R^2\,(1-\rho_{ST}^2)\; ,
		\end{equation}
		with $R^2=2.3$ and $5.99$ corresponding to $68.3 \%$  and
		$95 \% $  confidence levels (CLs) respectively.
		Our numerical analysis is performed with $\chi^2_{ST}$ at 95\% CL. 
		\item To further delimit the allowed parameter space, the \texttt{HiggsTools} package~\cite{Bahl:2022igd} is employed. This ensures that the allowed parameter regions align with the observed properties of the $125$~GeV Higgs boson  ( \texttt{HiggsSignals}~\cite{Bechtle:2013xfa,Bechtle:2014ewa,Bechtle:2020uwn,Bahl:2022igd}) and with the limits from searches for additional Higgs bosons at the LHC and at LEP ( \texttt{HiggsBounds}~\cite{Bechtle:2008jh,Bechtle:2011sb,Bechtle:2013wla,Bechtle:2020pkv,Bahl:2022igd}).
		\item[\textbullet]{\bf Flavour constraints}:  Flavour constraints are also implemented in our analysis. We used $B$-physics results, derived in \cite{Ouazghour:2023eqr} as well as  the experimental data at 2$\sigma$ \cite{HFLAV:2022pwe}  displayed in Table \ref{Tab2}.
	\end{itemize}
	{\renewcommand{\arraystretch}{1.5} 
		{\setlength{\tabcolsep}{0.1cm} 
			\begin{table*}[t]
				\centering
				\setlength{\tabcolsep}{7pt}
				\renewcommand{\arraystretch}{1.2} %
				\begin{tabular}{|l||c|c|}
					\hline
					Observable & Experimental result & 95\% C.L.\\\hline
					BR($\bar{B}\to X_{s}\gamma$)\cite{Ouazghour:2023eqr}&$(3.49\pm 0.19)\times10^{-4}$\cite{HFLAV:2022pwe}&$[3.11\times 10^{-4} , 3.87\times 10^{-4}]$\\\hline
				\end{tabular}
				\caption{Experimental result of flavor observable: $\bar{B}\to X_{s}\gamma$ at 95$\%$ C.L.}
				\label{Tab2}
			\end{table*}
			\vspace{6pt}
			\label{masssplit}	
			We perform scans within the allowed parameter space  by implementation of the theoretical and experimental constraints mentioned above.\\
			The following set of input parameters is used in the subsequent numerical analysis,

			\begin{align}
				\mathcal{P}_I = \big\{ &\alpha_1, \alpha_2, \alpha_3, m_{h_1}, m_{h_2}, \lambda_1, \lambda_3, \lambda_4, \lambda_6, \lambda_7, \lambda_8, \lambda_9, \bar{\lambda}_8, \bar{\lambda}_9,\nonumber\\
				& \mu_1, v_t, \tan\beta \big\}
				\label{parameters}
			\end{align}
			
			with,
			\begin{align}
				\begin{matrix}
					m_{h_1}= 125.09\,\text{GeV},\,\,m_{h_1}\leq m_{h_2}\leq m_{h_3}\leq 1\,\text{TeV},\,\,
					\\\frac{-\pi}{2}
					\leq \alpha_{1}\leq \frac{\pi}{2},\,\, -0.1\leq \alpha_{2,3}\leq 0.1\\
					0.5\leq \tan\beta\leq 120,\,\,-10^2\leq \mu_1\leq 10^2,\,\,0\leq v_t\leq 2 \,\text{GeV},\,\,\,\,\,\\-8\pi\leq \lambda_{i},\bar{\lambda_{i}}\leq 8\pi,
				\end{matrix}
				\label{input}
			\end{align}
			
in which $\tan\beta=v_2/v_1$.
			
We choose the lightest Higgs state $h_1$ to mimic  the observed SM-like Higgs boson at the LHC \cite{ATLAS:2012yve,CMS:2012qbp}  and set $m_{h_1}=125$ GeV. After carefully examining the model's parameter space in light of the aforementioned theoretical and experimental constraints, the resulting parameter space points will be used as input for \texttt{FormCalc}~\cite{Hahn:2001rv,Hahn:1998yk,Kublbeck:1990xc} to calculate the cross sections for each process.	
			
In Fig. \ref{correlation}, we display the correlation plots involving the masses of the charged Higgs and neutral Higgs bosons. The left panel shows the correlation between the pseudoscalar mass $m_{A_1}$ and the charged Higgs mass $m_{H_1^\pm}$, while the right panel illustrates the correlation between the heavier CP-even scalar $m_{h_2}$ and $m_{H_1^\pm}$. In both plots, the color code is used to represent the mass splittings $m_{H_1^\pm} - m_{A_1}$ and $m_{H_1^\pm} - m_{h_2}$, respectively, thus providing visual insight into how the degeneracy between these states evolves across the parameter space.

A clear pattern emerges from both panels, as the mass of the charged Higgs boson $m_{H_1^\pm}$ decreases, particularly in the region below 600 GeV, where the mass splitting between $m_{H_1^\pm}$ and the neutral Higgs bosons $A_1$ and $h_2$ tend to increase significantly. This shows that the regions with small values of the charged Higgs boson masses, are characterized by sizable mass splittings.

On the other hand, when $m_{H_1^\pm}$ exceeds approximately 600 GeV, the mass splittings between $H_1^\pm$ and both $A_1$ and $h_2$ are considerably reduced. In this heavier regime, the spectrum becomes more compressed, suggesting a near degeneracy among the heavy Higgs states. This behavior has important implications for collider phenomenology, as it influences both the decay modes and the kinematic distributions of final state particles.

\begin{figure*}[!ht]
	\begin{center}
		\includegraphics[scale=0.35]{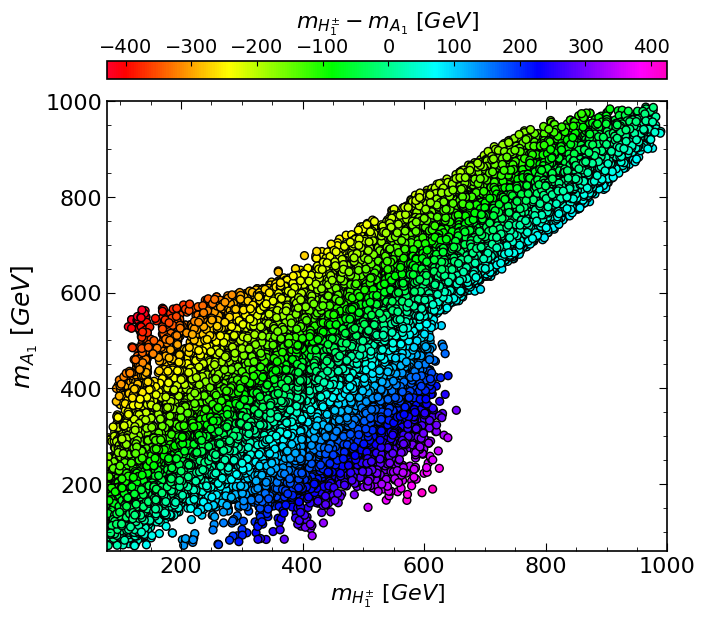}
		\includegraphics[scale=0.35]{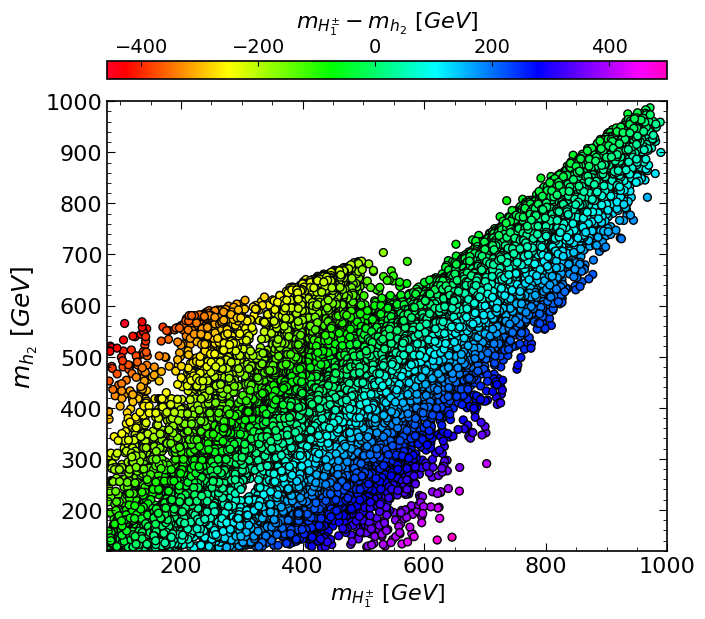}
	\end{center}
	\vspace{ -5mm}
	\caption{Correlations between $m_{A_1}$ and $m_{H_1^\pm}$ (left) and between $m_{h_2}$ and $m_{H_1^\pm}$ (right) after imposing the theoretical and experimental constraints. The color code indicates the splitting $m_{H_1^\pm} - m_{A_1}$ and $m_{H_1^\pm} - m_{h_2}$, respectively.}
	\label{correlation}
\end{figure*}
			
			
			%
			%
			\begin{itemize}
				\item {\color{blue}Results for $\mu^+ \mu^- \to H_1^+H_1^-$ and $\mu^+ \mu^- \to H_1^+H_2^-$}
			\end{itemize}

			\begin{figure*}[!htb]
			\begin{center}
				\includegraphics[scale=0.35]{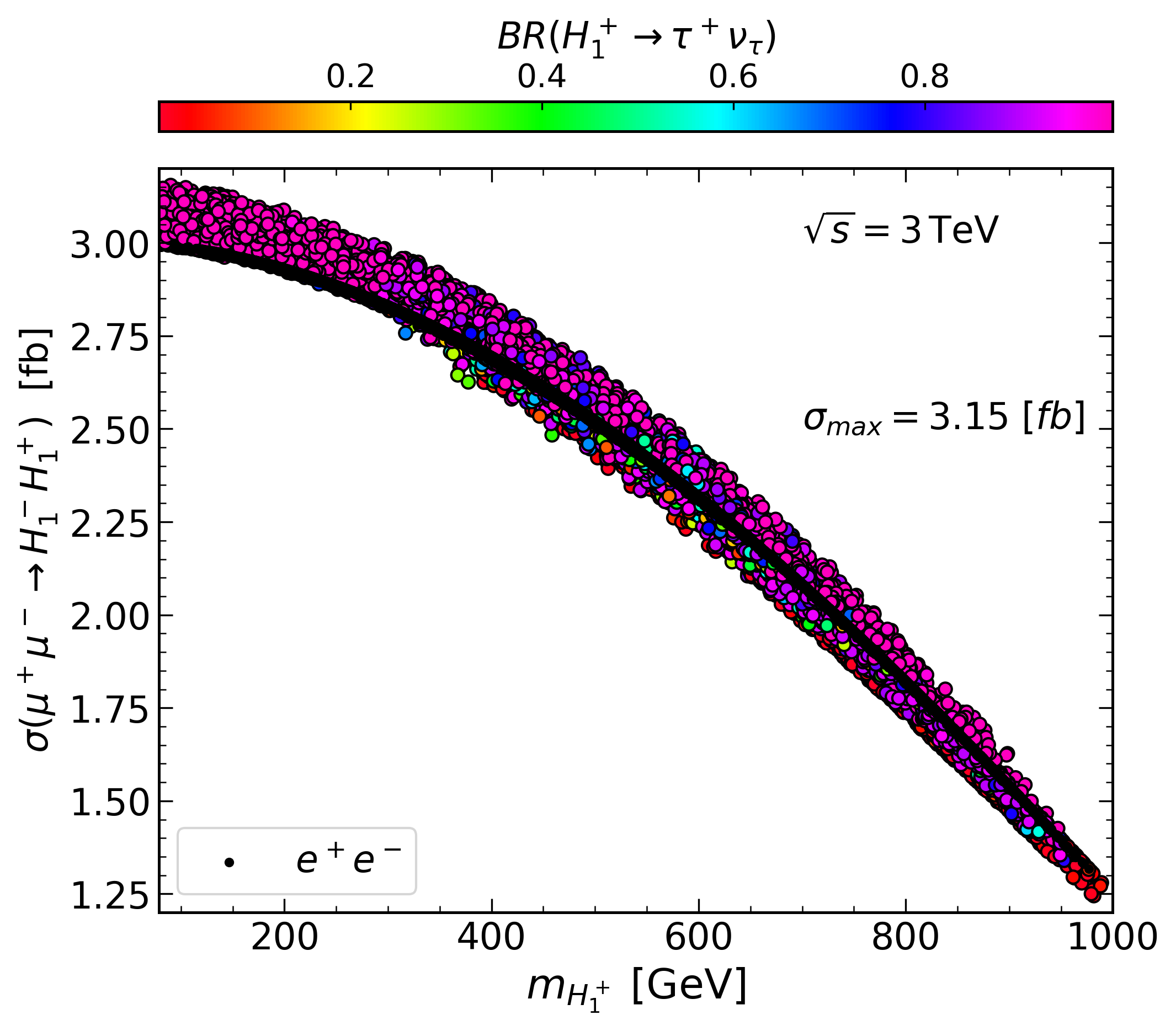}
				\includegraphics[scale=0.35]{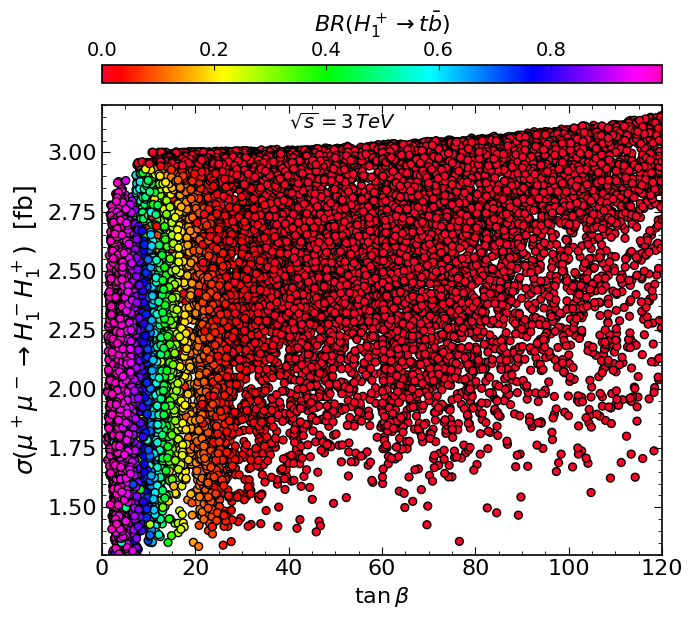}
				\includegraphics[scale=0.35]{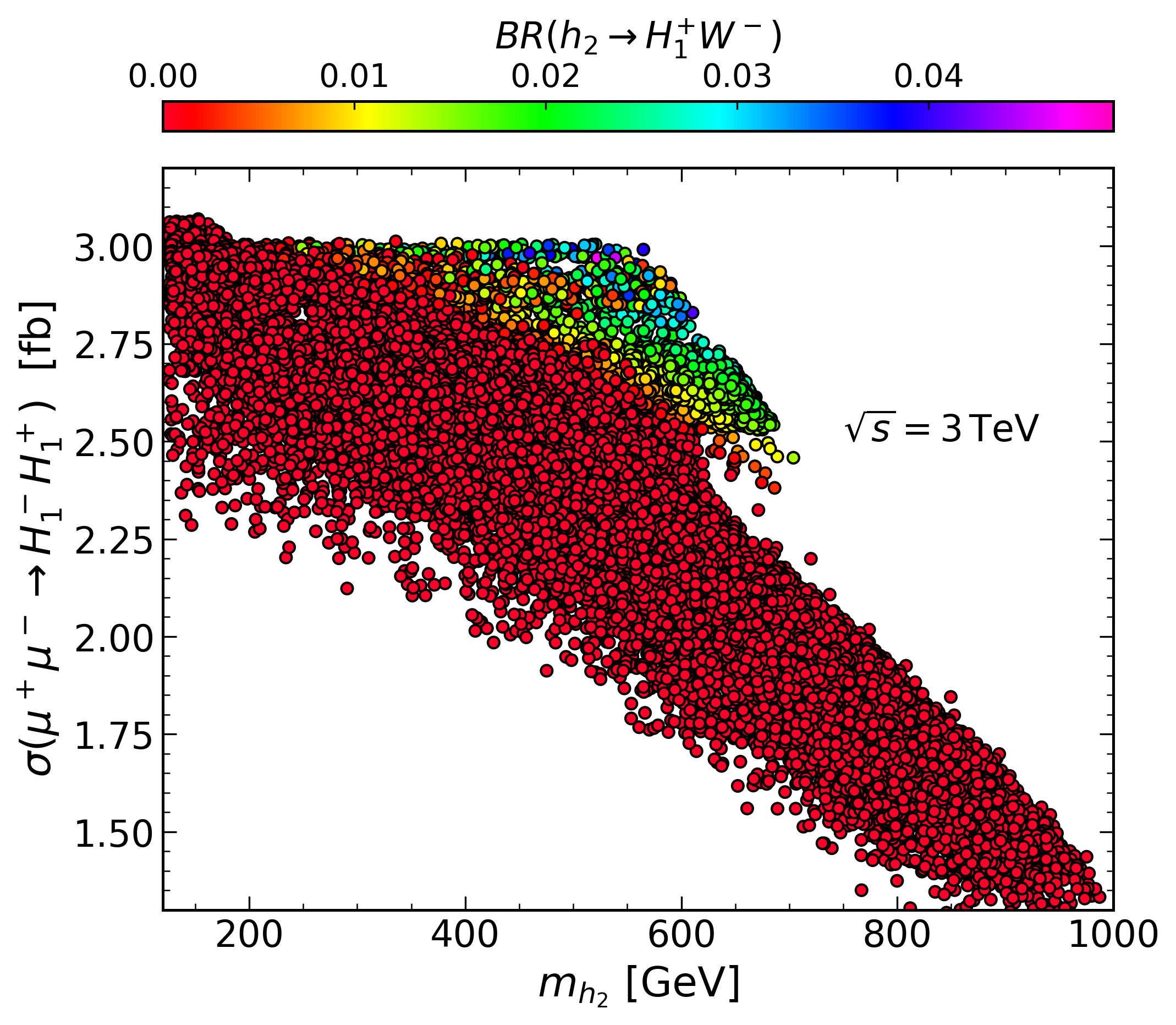}
				\includegraphics[scale=0.35]{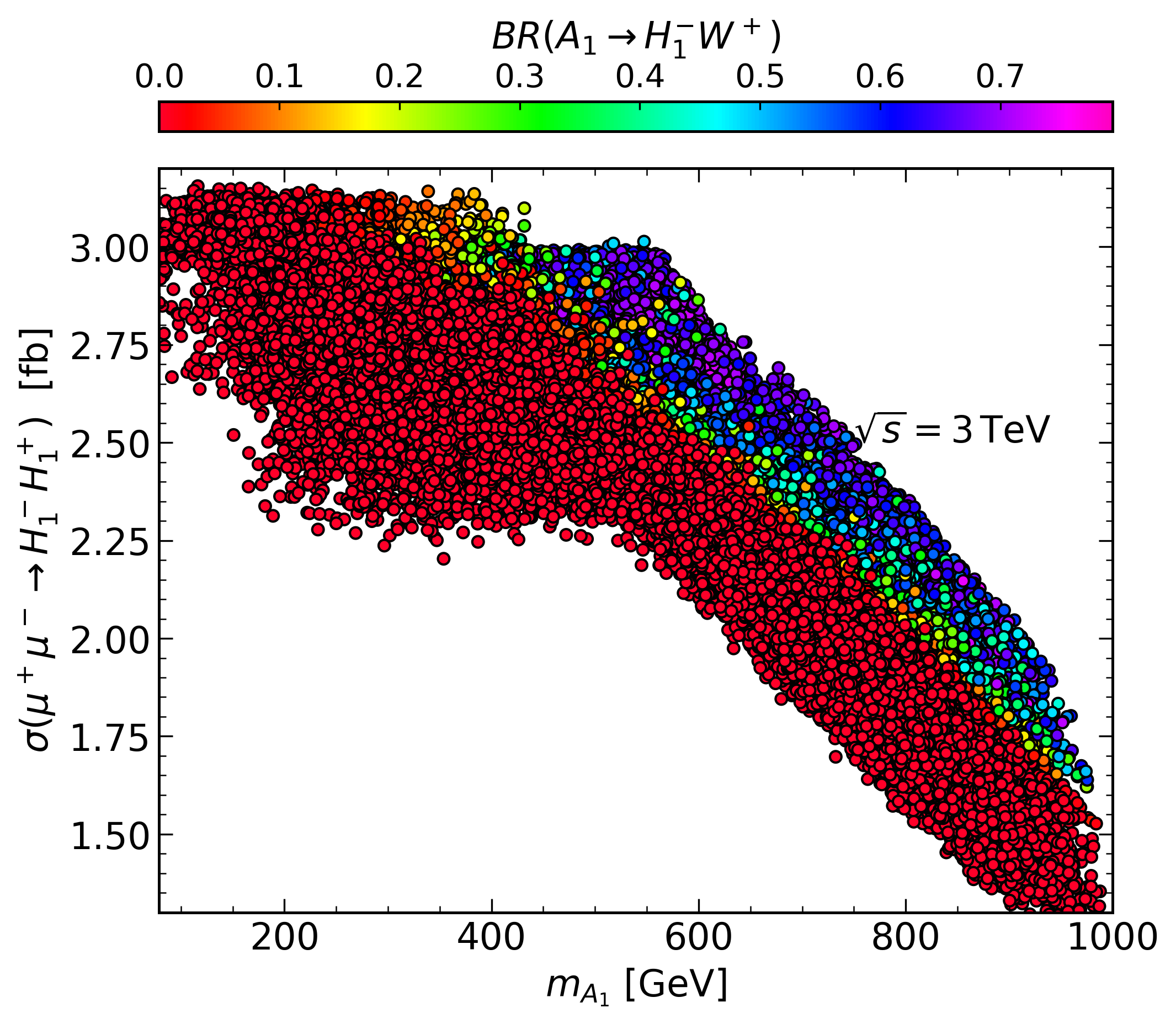}
		 \end{center}
			\vspace{ -5mm}
		\caption{
		\small
The upper panels we show the production cross section for $\mu^+\mu^- \to H_1^{\pm} H_1^{\mp}$ as a function of $m_{H_1^\pm}$ (left) and $\tan \beta$ (right) at $\sqrt{s}=3$ TeV. The color legend represent $Br(H_1^+ \to \tau^+\bar{\nu_\tau})$ and $Br(H_1^+ \to t\bar{b})$ respectively. The lower panels present the production cross section for $\mu^+\mu^- \to H_1^{\pm} H_1^{\mp}$ as a function of $m_{h_2}$ (left) and $m_{A_1}$ (right). The color coding presents $Br(h_2 \to H_1^-W^+)$ and $Br(A_1 \to H_1^-W^+)$ respectively . All the regions are consistent with theoretical and experimental constraints.}
				\label{results_Hp1Hm1}
			\end{figure*}	
The main results of our study on charged-Higgs pair production at a muon collider with a center-of-mass energy of 3 TeV are depicted in Fig. \ref{results_Hp1Hm1}. For this process, we first point out that the dominant contribution comes from the $s$-channels $\mu^+ \mu^- \to \gamma, Z \to H_1^+H_1^-$ diagrams (Fig.~\ref{diag:2mu2Hp1Hm1}$(d_{4,5})$). In addition the $s$-channels neutral-Higgs exchange $\mu^+ \mu^- \to h_1,h_2,h_3 \to H_1^+H_1^-$ (Fig \ref{diag:2mu2Hp1Hm1}$(d_{1,2,3})$) contributions, are generally smaller than $\mu^+ \mu^- \to \gamma, Z \to H_1^+H_1^-$. The $t$-channel contribution is also suppressed by $m_{\mu}^4$. Consequently, the cross section for charged-Higgs pair production is nearly identical to that obtained in $e^+e^- \to H_1^+ H_1^-$ at $e^+e^-$ colliders with the same $\sqrt{s}$. It is important to emphasize that the interference between the channels $\mu^+ \mu^- \to \gamma, Z \to H_1^+ H_1^-$ and $\mu^+ \mu^- \to h_1, h_2, h_3 \to H_1^+ H_1^-$ can be constructive, resulting in a total cross section that can match or even surpass $\sigma(\mu^+ \mu^- \to \gamma, Z \to H_1^+H_1^-)$=$\sigma(e^+e^- \to H_1^+ H_1^-)$.

The upper panels of Fig.~\ref{results_Hp1Hm1} illustrate the behavior of the production cross section for the charged Higgs pair production process $\mu^+ \mu^- \to H_1^+ H_1^-$ at a center-of-mass energy of $\sqrt{s} = 3$ TeV, as a function the model parameters, $m_{H_1^\pm}$ (shown in the left panel) and $\tan\beta$ (depicted in the right panel). The scatter plots reveal a clear inverse correlation between the cross section and the charged Higgs mass. Specifically, the production rate is significantly enhanced for lighter charged Higgs states, with the cross section reaching a maximum value of approximately 3.15 fb when $m_{H_1^\pm}$ is near the lower kinematic threshold.

On the other hand, the dependence of the cross section on $\tan\beta$ appears to be relatively weak, since across a wide range of $\tan\beta$ values, the variation in the cross section remains modest, suggesting that the pair production mechanism is largely insensitive to this parameter. In addition to the dependence on $m_{H_1^\pm}$ and $\tan\beta$, the lower panels of Fig.~\ref{results_Hp1Hm1} provide further insight into how the production cross section is influenced by the masses of the neutral Higgs bosons, namely $M_{h_2}$ and $M_{A_1}$. The plots clearly demonstrate that the cross section is substantially enhanced when both $M_{h_2}$ and $M_{A_1}$ are below approximately 570 GeV. 

Similarly, in Fig.~\ref{results_Hp1Hm2} we present our main results for the associated production of the lighter charged Higgs $H_1^{\pm}$ with the heavier one $H_2^{\pm}$, $\mu^+\mu^- \to H_1^{\pm} H_2^{\mp}$ at a muon collider with a center-of-mass energy of 3 TeV. The figure shows the production cross sections as functions of $m_{H_1^\pm}$ (left), $m_{H_2^\pm}$ (middle), and $\tan \beta$ (right). The color code indicates $\text{Br}(H_1^+ \to h_1 W^+)$ in the left and middle panels, and $\text{Br}(H_1^+ \to h_2 W^+)$ in the right panel. We clearly see that the cross section for this process is about an order of magnitude smaller than that of the pair production, reaching a maximum up to 0.17 fb.  In comparison, the cross section for $\mu^+ \mu^- \to H_1^{\pm} H_1^{\mp}$ is roughly 18 times larger, with $\sigma(\mu^+ \mu^- \to H_1^{\pm} H_1^{\mp}) \approx 18 \times \sigma(\mu^+ \mu^- \to H_1^{\pm} H_2^{\mp})$.
			
       \begin{figure*}[!htb]
		\centering
		\includegraphics[scale=0.33]{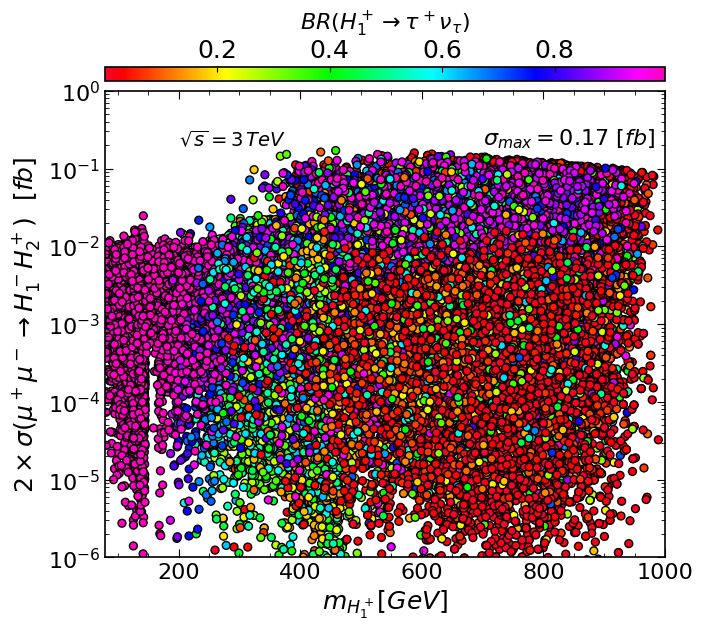}
		\includegraphics[scale=0.33]{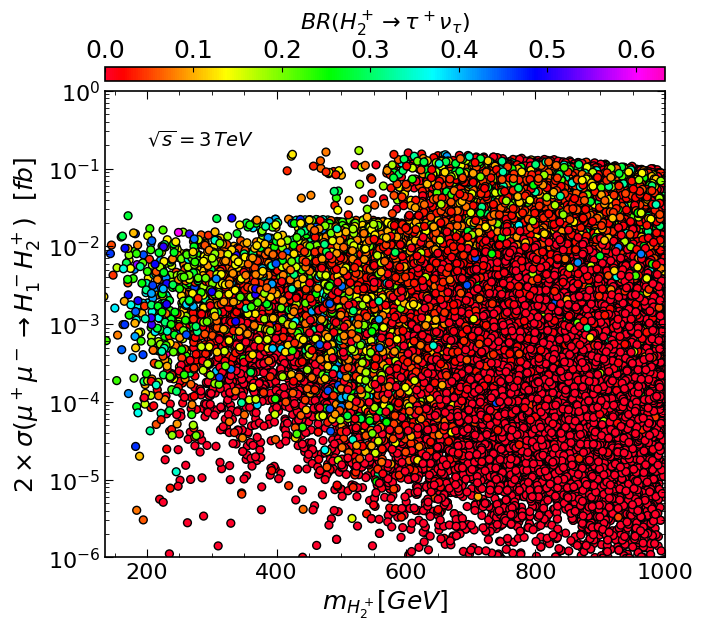}
		\includegraphics[scale=0.33]{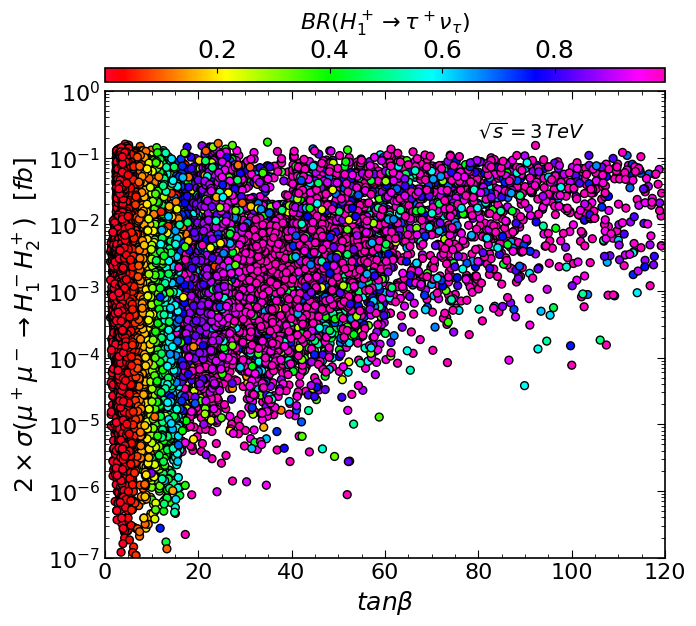}\\
		\caption{
		\small
		Production cross sections for $\mu^+\mu^- \to H_1^{\pm} H_2^{\mp}$ as a function of $m_{H_1^\pm}$ (left), $m_{H_2^\pm}$ (middle) and $\tan \beta$ (right) at $\sqrt{s}=3$ TeV. The color legend presents $Br(H_1^+ \to \tau^+\bar{\nu_\tau})$ (in left and right panels) and $Br(H_2^+ \to \tau^+\bar{\nu_\tau})$ (in middle panel) respectively. All the regions are consistent with theoretical and experimental constraints.}
		\label{results_Hp1Hm2}
	 \end{figure*}
			
	\begin{figure*}[!htb]
		\centering
	    \includegraphics[scale=0.32]{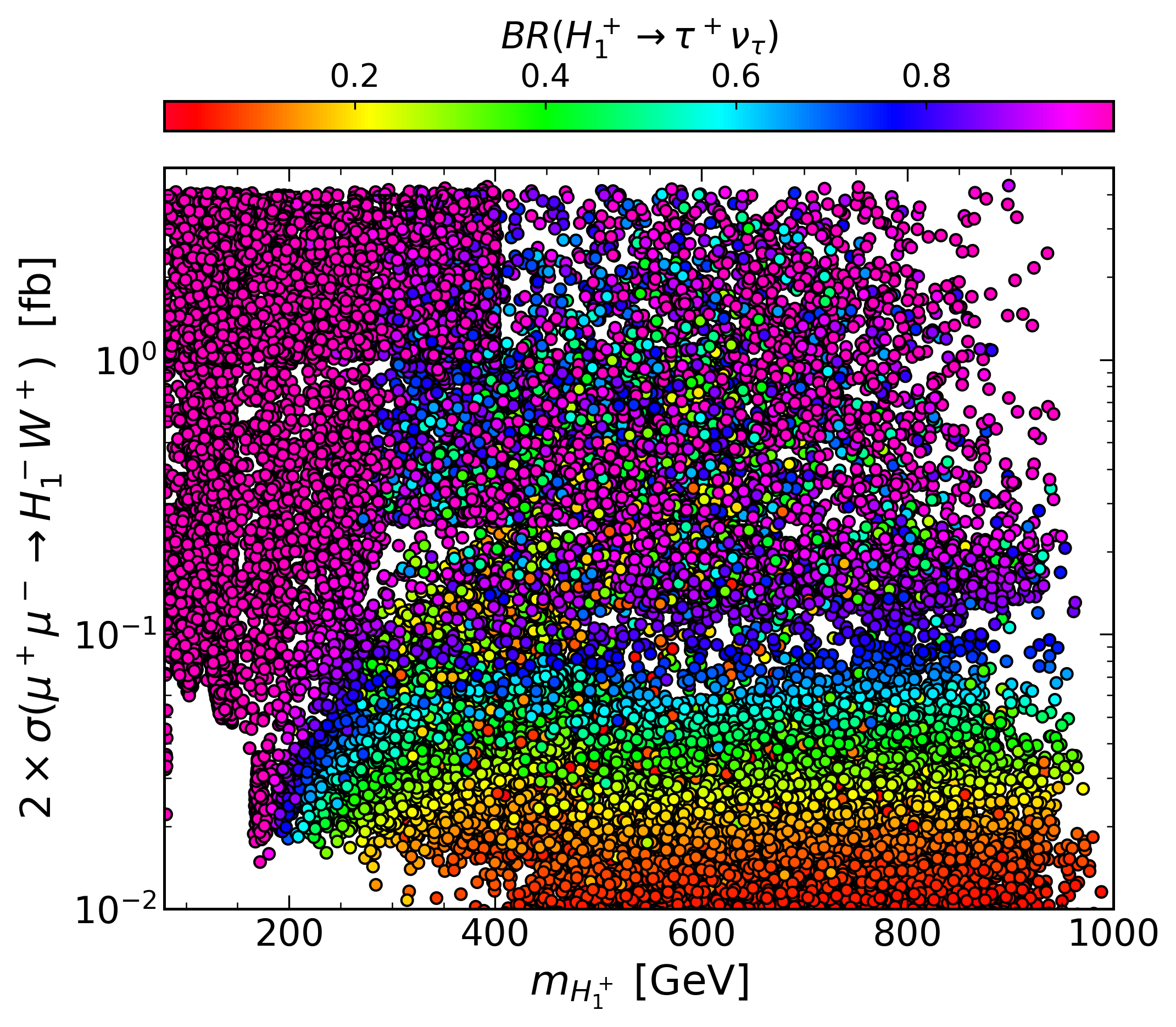}
		\includegraphics[scale=0.32]{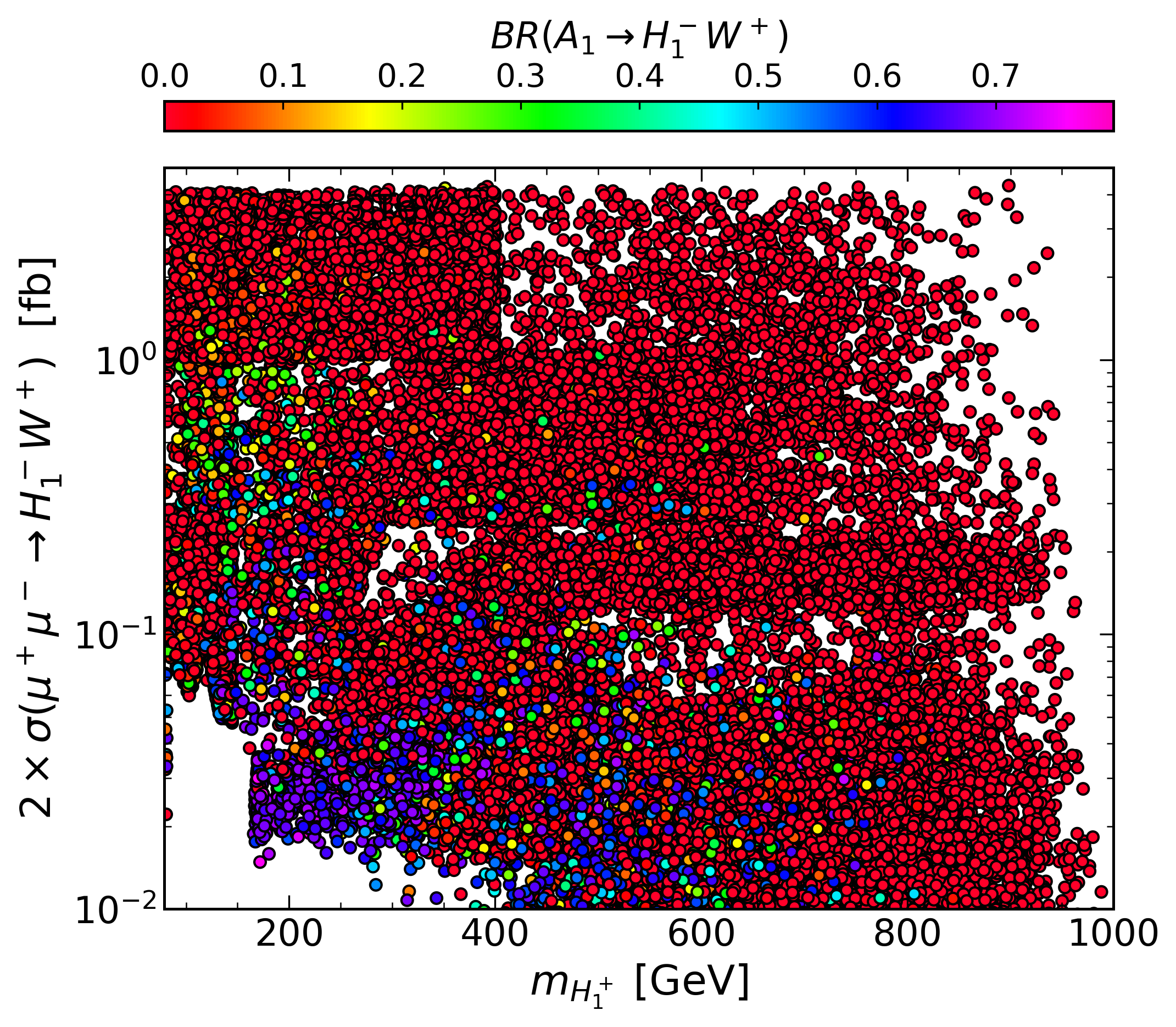}
		\includegraphics[scale=0.32]{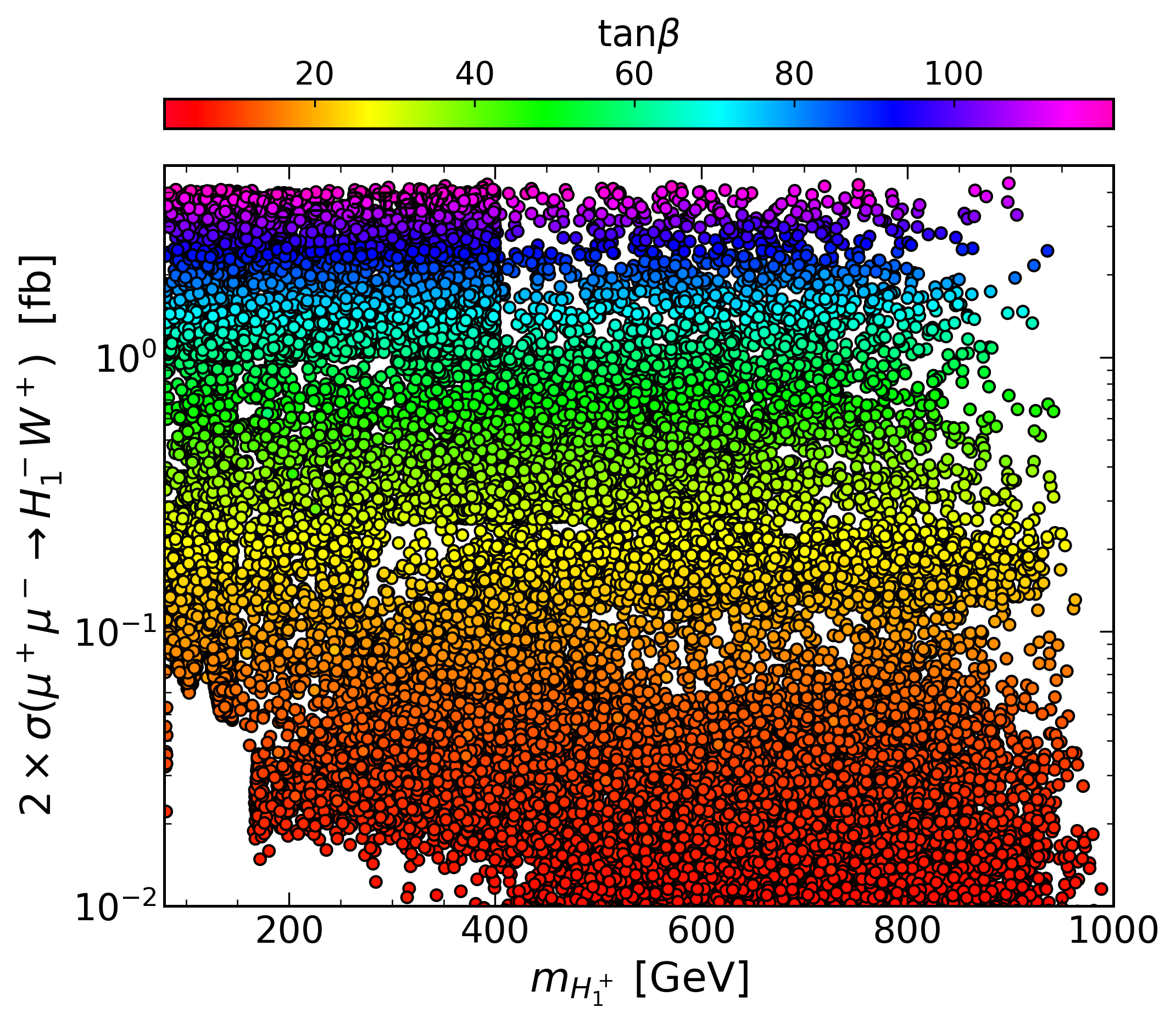}\\
		\caption{
		\small
		Production cross sections for $\mu^+\mu^- \to W^{\pm} H_1^{\mp}$ as a function of $m_{H_1^\pm}$ at $\sqrt{s}=3$ TeV. The color legend represents $Br(H_1^+ \to \tau^+\bar{\nu_\tau})$ in the (left panel) and $Br(A_1^+ \to H_1^{+}W^-)$ (middle panel) and tan$\beta$ (right panel). All the regions are consistent with theoretical and experimental constraints.}
		\label{results_wh}
	\end{figure*}

			\begin{itemize}
				{\color{blue}\item Results for  $\mu^+\mu^- \to H_1^\pm W^\mp$}
			\end{itemize}
Unlike the pair production process $\mu^+\mu^- \to H_1^\pm H_1^\mp$ and $\mu^+\mu^- \to H_1^\pm H_2^\mp$, the current process is not sensitive to pure scalar couplings due to the presence of a $W$ boson in the final state. Moreover, the $s$-channel diagram for $\mu^+\mu^- \to h_i, A_j \to H_1^\pm W^\mp$ (where $i=1,2,3$ and $j=1,2$) can be either suppressed or enhanced based on the values of couplings $h_i W^\pm H_1^\mp$ and $A_j W^\pm H_1^\mp$. In contrast to the process $\mu^+\mu^- \to H_1^\pm H_1^\mp$, the $t$-channel contribution for $\mu^+\mu^- \to W^\pm H_1^\mp$ with neutrino exchange is not negligible, although it remains smaller compared to the $s$-channel contribution involving neutral Higgs exchanges. In the case of $\mu^+\mu^- \to H_1^\pm H_1^\mp$, the $t$-channel amplitude is proportional to $m_{\mu}^2$, while for $\mu^+\mu^- \to W^\pm H_1^\mp$ the $t$-channel amplitude has only one $m_{\mu}$ suppression which could be overcome with the large $\tan\beta$ value. Therefore, both the $t$- and $s$-channel contributions are proportional to $\tan\beta$ in the large $\tan\beta$ limit, since both $h_i\mu^+\mu^- \propto 1/c_{\beta}= \sqrt{1+\tan\beta^2}\approx \tan\beta$. In the large $\tan\beta$ limit, the amplitudes of all diagrams are proportional to $\tan\beta$ and it can take large values in Type-III, therefore we would expect enhancement for large $\tan\beta$ for Type III.

In Fig.~\ref{results_wh}, we show the production cross section for the process $\mu^+ \mu^- \to H_1^\pm W^\mp$ at a future high-energy muon collider operating at a center-of-mass energy of $\sqrt{s} = 3$ TeV. The results are displayed as a function of the mass of the charged Higgs boson $H_1^\pm$. The color coding in the plots indicates information about the relevant decay branching ratios and parameters, specifically, the left is color coded according to the branching ratio $\text{Br}(H_1^+ \to \tau^+ \bar{\nu}_\tau)$, the middle according to $\text{Br}(A_1 \to H_1^\pm W^\mp)$ while the right panel highlights the values of tan$\beta$.

From the plots, we clearly see that the cross section for this process can be of the same order of magnitude as that of the charged Higgs pair production process $\mu^+ \mu^- \to H_1^+ H_1^-$, indicating that both channels may play a comparably significant role in probing charged Higgs bosons at future muon colliders. However, a distinctive and noteworthy feature of the $\mu^+ \mu^- \to H_1^\pm W^\mp$ process is its noticeable sensitivity to the parameter $\tan\beta$. Unlike the pair production channel, which is typically less affected by variations in $\tan\beta$, the associated production cross section increases significantly with larger values of $\tan\beta$, reaching a maximum of approximately 4.12 fb in the considered parameter space exceeding that of the pair production.

Moreover, an inverse correlation is observed between the magnitude of the cross section and the branching ratio $\text{Br}(A_1 \to H_1^\pm W^\mp)$, as seen in the middle panel. Specifically, regions of parameter space characterized by a suppressed decay of the pseudoscalar $A_1$ into $H_1^{\pm}$ and a $W^{\mp}$ bosons tend to exhibit enhanced production cross sections for the $\mu^+ \mu^- \to H_1^\pm W^\mp$ process.

\section{Signal-Background analysis
			}
\label{section5}
\subsection{Monte Carlo Toolchain}
\label{section3subsec4}
			
In this study we examine the potential of a future muon collider operating at $\sqrt{s} = 3$ TeV, focusing on final states involving a pair of tau-leptons accompanied by missing energy. Specifically, we explore the processes $\mu^{+} \mu^{-} \rightarrow H_1^{+} H_1^{-} \rightarrow \tau^+ \nu \tau^- \nu$ and $\mu^{+} \mu^{-} \rightarrow H_1^{+} W^{-} \rightarrow \tau^+ \nu \tau^- \nu$. For each studied channel, we adopt the benchmark scenarios outlined in Table~\ref{Bp1}. 
			
Signal events are simulated by generating the parton-level processes using \texttt{MadGraph5\_aMC\_v3.4.2}  \cite{Alwall:2014hca}. The generated events are then passed through \texttt{Pythia-8.20} \cite{Sjostrand:2007gs} to account for fragmentation and parton showering. Finally, the events undergo detector simulation using \texttt{Delphes-3.4.5} \cite{deFavereau:2013fsa}, where the muon collider Detector TARGET model is implemented. Jets are reconstructed using the anti-$k_t$ algorithm \cite{Cacciari:2008gp} implemented in \texttt{Delphes}, with a jet radius parameter of $R = 0.5$. At the \texttt{Delphes} simulation level, a jet candidate is required to meet a minimum transverse momentum threshold of $p_T > 20$ GeV to qualify as a b-jet. Additionally, a b-tagging efficiency of approximately 70$\%$ is applied, along with mistag rates for charm and light-quark jets being misidentified as b-jets. These rates are treated as functions of the jet pseudorapidity and energy.
			
The significance of the signal is evaluated using the median significance approach~\cite{Cowan:2010js}. The discovery significance ($\mathcal{Z}_\mathrm{disc}$) was calculated using the following formulas :
\begin{widetext}
	\begin{eqnarray}
		\mathcal{Z}_\mathrm{disc} &=& \sqrt{2\left[\left(s+b\right)\ln\left(\frac{\left(s+b\right)\left(1+\delta^2b\right)}{b+\delta^2b\left(s+b\right)}\right)-\frac{1}{\delta^2}\ln\left(1+\delta^2\frac{s}{1+\delta^2b}\right)\right]},
	\end{eqnarray}
\end{widetext}			
			\begin{table*}[t]
				\setlength{\tabcolsep}{2pt}
				\renewcommand{\arraystretch}{1.2}
				\centering
				\begin{tabular}{|c|c|c|c|c|c|c|c|c|c|c|c|c|c|c|c|c|c|c|c|c}       
					\hline \hline 
					&signal& $m_{h_1}$  & $m_{h_2}$ & $\lambda_1$ & $\lambda_3$ & $\lambda_4$  &  $\lambda_6$& $\lambda_7$ & $\lambda_8$ & $\lambda_9$ & $\bar{\lambda_8}$ & $\bar{\lambda_9}$ & $\tan \beta$ & $\alpha_{1}$& $\alpha_{2}$ & $\alpha_{3}$ & $v_t$ \\ \hline
					
					\text{BP1}&$ \tau^+ \nu \tau^- \nu$ & 125.09 & 152.82& 0.70 & -0.14 & 0.76 & 0.48 & 0.48 & 0.71 & -0.036 & 2.97 & -0.19 & 89.06 & 1.561 & $\approx$0& $\approx$0 &0.18\\ \hline
					\text{BP2}&$ \tau^+ \nu \tau^- \nu$ & 125.09 & 181.94& 3.03 & 0.77 &  3.28 & 1.50 & 0.01 & -0.018 &0.40 & 1.55 & 1.60 & 9.96 & -1.47 & -0.025& -0.62 &0.057\\ \hline

					\text{BP3}&$ \tau^+ \nu \tau^- \nu$ & 125.09 & 673.49& 0.29 & 4.08 &  -2.73 & 4.28 & 5.31 & 0.82 &-1.49 & 0.65 & 0.71 & 119.29 & 1.563 & $\approx$0& $\approx$0 &0.19\\ \hline
					\hline \hline
				\end{tabular}
				\caption{The description of our BPs.}\label{Bp1}
			\end{table*}

			\subsection{{\color{blue}$\mu^+\mu^- \to H_1^+H_1^-$}}
In this subsection, we focus on pair production of charged Higgs at future muon collider with the final state $\tau^{+} \nu \tau^{-} \nu $.
Muon colliders provide a superior platform compared to the LHC, offering higher production rates, significantly reduced backgrounds, and precise energy control\footnote{Charged Higgs boson pair production at the LHC suffers from low cross sections, overwhelming backgrounds, and energy limitations.}. These features make the process $\mu^+\mu^- \to H_1^+H_1^-$ not only feasible but also a promising focus for discovery and precision measurements at muon colliders.
{\color{blue}\[ \mu^{+} \mu^{-} \rightarrow H_1^{+} H_1^{-} \rightarrow  \tau^+ \nu_{\tau} \tau^- \nu_{\tau} \]}
To search for signals amidst the SM background, we highlight the primary SM backgrounds, this includes top-pair production ($t\bar t$), diboson production ($VV$ with $WW$ and $ZZ$), $Zjj$, and $Wjj$, as outlined in Table~\ref{backgrounds}.
			\begin{table*}[t]
				\centering
				\begin{tabular}{|c|c|}
					\hline
					\textbf{Process} & \textbf{Description} \\
					\hline
					$\mu^{+} \mu^{-} \rightarrow t\bar{t}$ & Top-pair production \\
					\hline
					$\mu^{+} \mu^{-} \rightarrow VV$ & Diboson production ($WW$, $ZZ$) \\
					\hline
					$\mu^{+} \mu^{-} \rightarrow Z/\gamma \ jj$ & $Z \rightarrow \tau \tau$ and $Z \rightarrow \nu \nu$ \\
					\hline
					$\mu^{+} \mu^{-} \rightarrow Wjj$ & One $\tau$ from $W$ decay, the other $\tau$ from a jet misidentified as $\tau$ \\
					\hline
				\end{tabular}
				\caption{The main SM backgrounds.}
				\label{backgrounds}
			\end{table*}
			
To ensure that the events at the parton-level meet the necessary criteria, we impose $p_T^{j} > 25$ and $|\eta_{j}| < 2.5$ on jets. At the Delphes level we apply the $\tau$-tagging efficiencies and the mistag rates of a light jet as a $\tau$, $P_{\tau \rightarrow \tau}$=0.85 and $P_{j \rightarrow \tau }$=0.02, respectively. In addition, we enforce the charged lepton identification and standard photon isolation criteria, with the requirement,
	\begin{center}
				$I(P)= \frac{1}{p_{T}^{P}} \Sigma p_{T_{i}} < 0.01$   :
			\end{center} 
			\begin{figure*}[!htb]
				\centering	
				\includegraphics[scale=0.39]{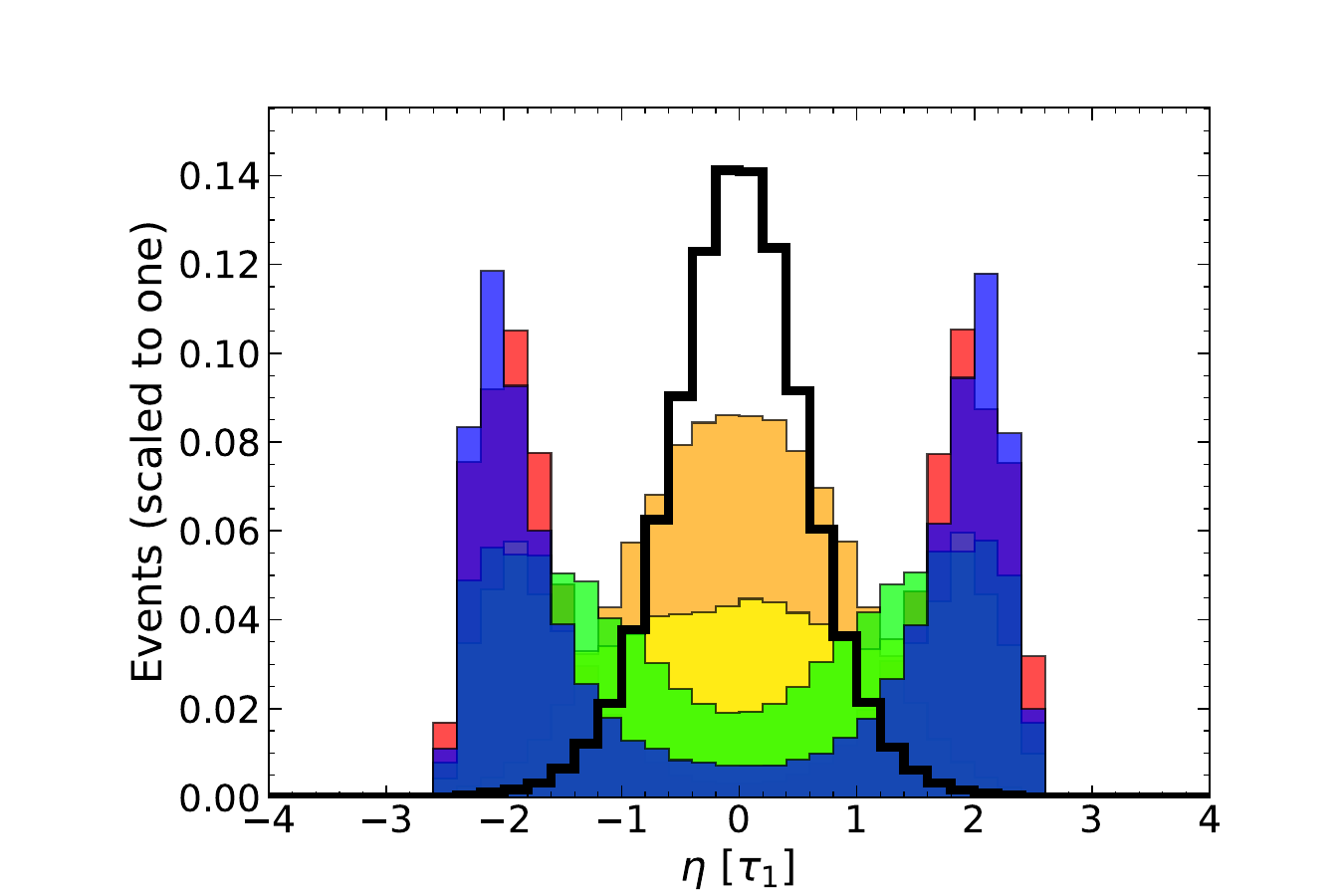}
				\includegraphics[scale=0.39]{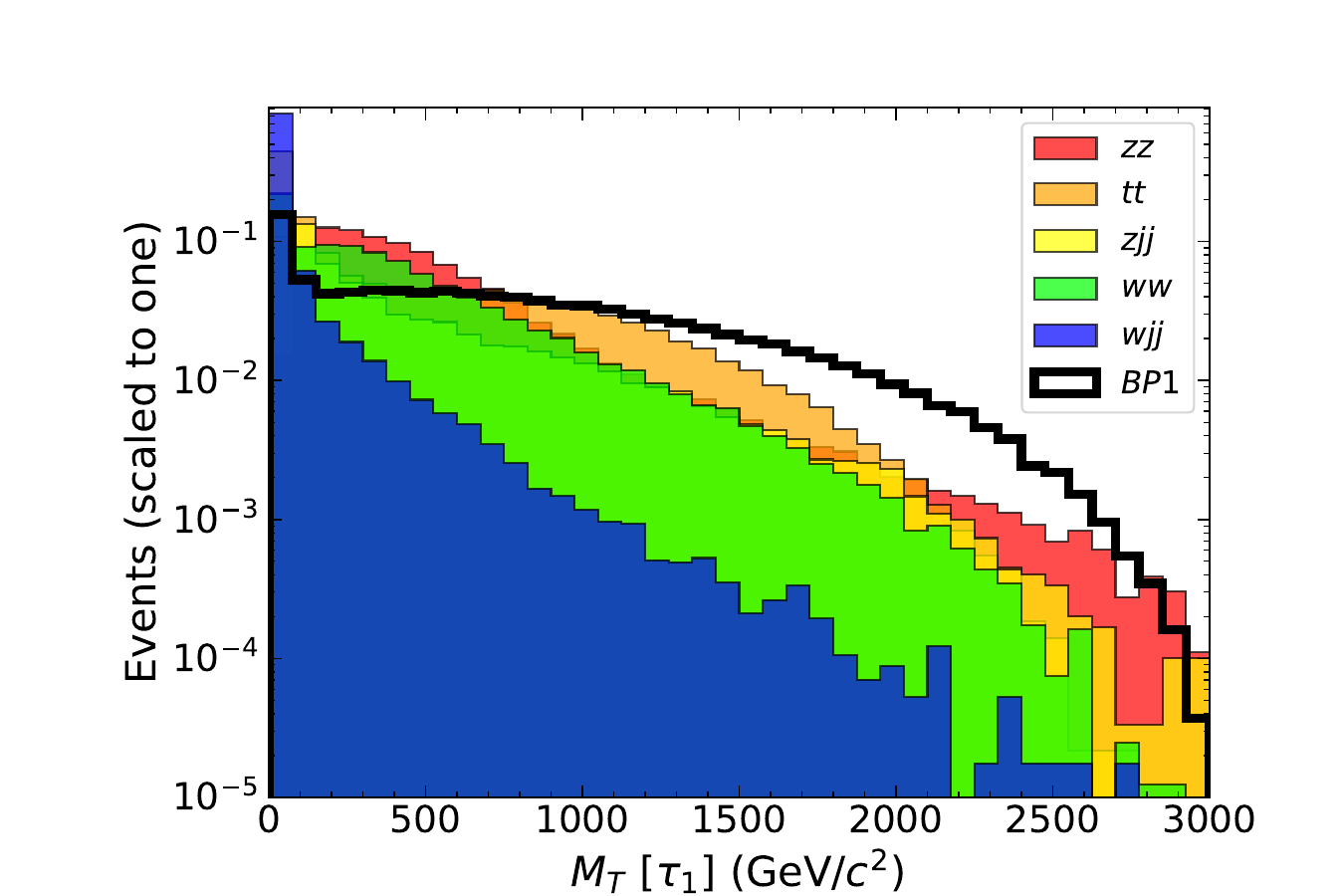}
				\caption{Kinematic distributions for the $\tau^+ \nu_{\tau} \tau^- \bar{\nu_{\tau}}$ final state at a $\sqrt{s}=3$ TeV muon collider : pseudorapidity $\eta(\tau_1)$ of the leading tau lepton (left) and transverse mass $M_T(\tau_1)$ (right).}
				\label{p1}
   \end{figure*}
			
			\begin{table*}[t]
				\centering
				\renewcommand{\arraystretch}{1.3}
				\setlength{\tabcolsep}{45pt}
				\begin{adjustbox}{max width=\textwidth}
					\begin{tabular}{l c} 
						\hline \hline
						\textbf{Cuts}  & \textbf{Definition} \\  
						\hline \hline	
						\textbf{Trigger} & $N(b) \leq 1$\quad and \quad $N(j) \leq 1$ \\  	
						\hline
						\textbf{Cut-1} & $-0.8 < \eta [\tau_{1}] < 0.8$  \\   
						\hline
						\textbf{Cut-2} & $M_T[\tau_{1}] > 200$ GeV \\
						\hline \hline
					\end{tabular}
				\end{adjustbox}		
				\caption{Selection criteria applied in the signal-background analysis of the process 
					$\mu^+ \mu^- \to H_1^+ H_1^- \to \tau^+ \nu_{\tau} \tau^- \nu_{\tau}$ 
					at $\sqrt{s} = 3$ TeV.}
				\label{selection_cuts_Hp1Hm1}
			\end{table*}
\begin{table*}[ht!]
\centering
\setlength{\tabcolsep}{8pt}
\renewcommand{\arraystretch}{1.2}
\begin{tabular}{l c c c c c c}
\hline\hline
\multirow{2}{*}{\textbf{Cuts}} & \textbf{Signal} & \multicolumn{5}{c}{\textbf{Backgrounds}} \\ 
\cline{2-2} \cline{3-7} 
& BP1 & $WW$ & $ZZ$ & $t\bar t$ & $Wjj$ & $Z/\gamma jj$ \\
\hline\hline
Basic cut  & 3.143 & 1.54 & 0.0147 & 0.21 & 7.04 & 0.74 \\
Tagger  & 3.111 & 1.508 & 0.01465 & 0.0901 & 2.76 & 0.16 \\
Cut-1  & 2.115 & 0.234 & 0.000417 & 0.0639 & 0.0564 & 0.0133 \\
Cut-2  & 1.634 & 0.166 & 0.000382 & 0.0517 & 0.0162 & 0.00949 \\
\hline
\textbf{Total efficiencies} & \textbf{52\%} & \textbf{10.8\%} & \textbf{2.6\%} & \textbf{25.8\%} & \textbf{0.231\%} & \textbf{1.28\%} \\
					\hline\hline
\end{tabular}
\caption{Cut flow of the cross sections (in fb) for the signal and SM backgrounds at $\sqrt{s} = 3$ TeV muon collider using the benchmark point (BP1).}
\label{cutft_HP1Hm1}
\end{table*}

We illustrates in Fig.~\ref{p1} the pseudorapidity of the leading tau $\eta\ (\tau_1)$ and transverse mass $M_{T}(\tau_{1})$. These distributions are shown for the signal benchmark point BP1 and various SM backgrounds at the 3 TeV muon collider.	
In order to increase the significance of the signal, we have established a cut-flow based on the behavior of the kinematic distributions, as presented in Table~\ref{cutft_HP1Hm1}.
To enhance signal-background discrimination, we impose constraints on the number of b-quarks and jets, requiring N(b)$\leq 1$ and N(j)$\leq 1$, which is vital in discriminating the signal from the background.
Under this cut 44.9$\%$ of $t\bar t$, 21.6$\%$ of $z jj$ and 39.4$\%$ of Wjj backgrounds events survived, while the signal remains unaffected. Next we apply an additional selection cut $-0.8< \eta [\tau_{1}]<0.8$, which effectively suppresses a significant portion of the background, particularly eliminating most of the $Wjj$, $ZZ$, $WW$ and $Zjj$ events, while retaining approximately 70.9$\%$ of the $t\bar{t}$ events and 68$\%$ of the signal events. We then apply an additional cut on the invariant mass, requiring $M_{T}[\tau_{1}] > 200$ GeV. This effectively removes $71.3\%$ of $Wjj$ events while 77.3$\%$ of the signal survives. The selection cuts imposed are listed in Table~\ref{selection_cuts_Hp1Hm1}.
\begin{table*}[!htb]
	\centering
	\includegraphics[scale=0.70]{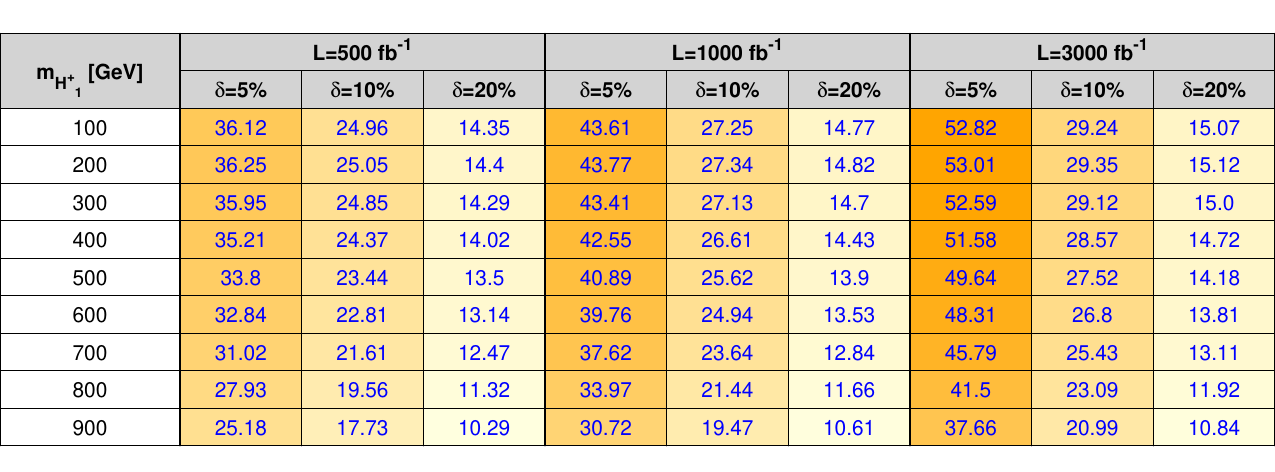}	
	\caption{The significance for the process $\mu^+ \mu^- \to H_1^\pm H_1^\mp \to \tau^+ \nu_{\tau} \tau^- \nu_{\tau}$ is presented as a function of the charged Higgs mass at a 3 TeV muon collider, considering integrated luminosities of 500 fb$^{-1}$, 1000 fb$^{-1}$, and 3000 fb$^{-1}$ and varying systematic uncertainties $\delta$, using our benchmark points presented in Table~\ref{Hp1Hm1_BPs_table}.}
	\label{SigHp1Hm1}
\end{table*}

\begin{table*}[t]
	\scriptsize  
	\setlength{\tabcolsep}{1.6pt}  
	\renewcommand{\arraystretch}{1.3}  
	\centering
	\begin{tabular}{|c|cccccccccccccccccc|}
		\hline \hline
		& $m_{h_1}$ & $m_{h_2}$ & $\lambda_1$ & $\lambda_3$ & $\lambda_4$ & $\lambda_6$ & $\lambda_7$ & $\lambda_8$ & $\lambda_9$ & $\bar{\lambda}_8$ & $\bar{\lambda}_9$ & $\tan\beta$ & $\alpha_1$ & $\alpha_2$ & $\alpha_3$ & $v_t$ & $\mu_1$ \\
		\hline
		BP1' & 125.09 & 146.32 & 0.863 & -0.118 & 0.394 & 0.836 & 0.157 & 0.422 & 0.00944 & 1.59 & 0.798 & 43.97 & 1.55 & -5.25e-4 & 0.0964 & 0.495 & 17.04 \\
		BP2' & 125.09 & 240.47 & 0.0755 & 0.841 & 0.396 & 2.03 & 5.41 & 1.87 & 1.66 & 1.13 & -0.768 & 59.45 & -1.56 & -2.93e-4 & -0.0464 & 0.324 & 56.06 \\
		BP3' & 125.09 & 290.17 & 0.381 & 0.687 & -0.186 & 6.30 & 2.66 & -2.58 & 4.46 & -0.544 & 1.79 & 46.98 & 1.55 & -5.29e-4 & -0.0533 & 0.179 & -69.86 \\
		BP4' & 125.09 & 396.44 & 0.00333 & 0.711 & -0.223 & 2.88 & 3.18 & 3.64 & 3.36 & 1.05 & -0.0548 & 52.66 & 1.55 & -2.97e-4 & -0.00464 & 0.0784 & 94.91 \\
		BP5' & 125.09 & 419.76 & 0.543 & 2.66 & -0.541 & 1.67 & 2.68 & 1.25 & 0.225 & 0.518 & 0.679 & 70.81 & 1.56 & 2.95e-4 & 0.997 & 0.0957 & -31.00 \\
		BP6' & 125.09 & 541.61 & 0.103 & 2.70 & -0.114 & 1.13 & 4.40 & 0.351 & 2.74 & 0.136 & -0.00141 & 62.48 & 1.56 & -8.39e-4 & 0.0735 & 0.379 & -26.78 \\
		BP7' & 125.09 & 623.47 & 0.524 & 3.96 & -2.54 & 1.58 & 2.98 & 4.68 & -1.83 & 0.302 & 1.32 & 105.84 & 1.56 & 1.11e-4 & 0.998 & 0.0592 & 55.95 \\
		BP8' & 125.09 & 743.38 & 0.233 & 3.16 & -1.17 & 6.12 & 4.74 & -1.07 & 0.163 & 1.15 & -0.944 & 94.11 & 1.56 & -3.96e-4 & 0.99 & 0.272 & 5.19 \\
		BP9' & 125.09 & 875.32 & 0.924 & 2.86 & -0.406 & 5.12 & 3.69 & 1.77 & -0.477 & 0.265 & 0.378 & 108.94 & 1.56 & -3.4e-5 & 0.98 & 0.0986 & -39.06 \\
		\hline \hline
	\end{tabular}
	\caption{Benchmark points used in Table~\ref{SigHp1Hm1}.}
	\label{Hp1Hm1_BPs_table}
\end{table*}

In order to provide a more complete overview of the the discovery prospects for charged Higgs boson $H_1^{\pm}$, we extend the analysis to cover a wide mass range from 100 GeV to 900 GeV. Based on the behavior of the kinematic distributions of the previously examined benchmark point (BP1), we present the discovery significance ($\mathcal{Z}_\mathrm{disc}$) for each mass point\footnote{It is important to note that, for each mass point, the parameter point yielding the highest possible cross section is considered. See Table~\ref{Hp1Hm1_BPs_table}.} at a center-of-mass energy of $\sqrt s$ = 3 TeV, considering
integrated luminosities of 500 fb$^{-1}$, 1000 fb$^{-1}$, and 3000 fb$^{-1}$, as well as systematic uncertainties ($\delta$)
of 5 $\%$, 10$\%$, and 20 $\%$. All evaluated parameter points satisfy both theoretical and experimental constraints, ensuring the reliability of the explored parameter space. The discovery significances obtained for the entire mass range under the various luminosity scenarios are summarized in Table~\ref{SigHp1Hm1}.

The results demonstrate that even at the lower bound of projected luminosity, $\mathcal{L} = 500$ fb$^{-1}$, the experiment retains substantial sensitivity across a wide range of the charged Higgs boson mass $m_{H_1^{+}}$. In particular, the significance consistently exceeds the $5\sigma$ discovery threshold ($\mathcal{Z}_\mathrm{disc} \geq 5$) throughout the investigated mass spectrum and for the different considered systematic uncertainties ($\delta$) of 5 $\%$, 10$\%$, and 20 $\%$, suggesting that a moderate data collection could be feasible for  a potential observation of the charged Higgs boson. However, important observations can be made. 

First, the significance decrease with the increase of the systematic uncertainties ($\delta$). For example, at $\delta = 20\%$, the significance is noticeably reduced compared to the case with $\delta = 5\%$, about 60$\%$ decrease, underscoring the necessity of a strong control over systematic sources of uncertainty for a robust discovery prospects. 

Secondly, for each fixed value of $\delta$ the significance decrease as the mass of the charge Higgs get larger. This is a direct consequence of the reduced production cross section at higher masses. Moreover, as expected, we observe that increasing the integrated luminosity leads to a notable improvement in the statistical significance.
\begin{figure}[!htb]
	\centering
	\includegraphics[scale=0.40]{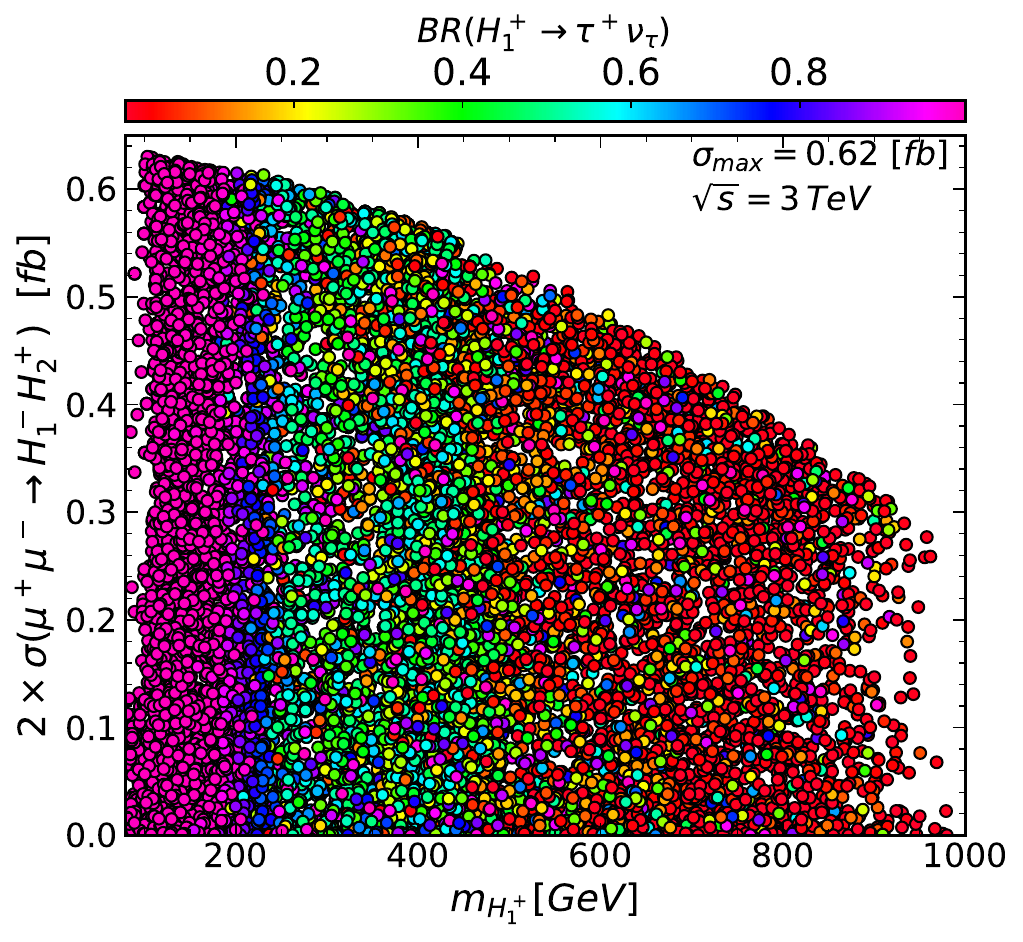}\\
	\caption{
		\small
		Production cross sections for $\mu^+\mu^- \to H_1^{\pm} H_2^{\mp}$ as a function of $m_{H_1^\pm}$ at $\sqrt{s}=3$ TeV. The color legend presents $Br(H_1^+ \to \tau^+\bar{\nu_\tau})$. All the regions are consistent with all the theoretical and experimental constraints except the EWPO constraints.}
	\label{results_Hp1Hm2_NO_STU}
\end{figure}	
\begin{table*}[!htb]
	\centering
	\includegraphics[scale=0.70]{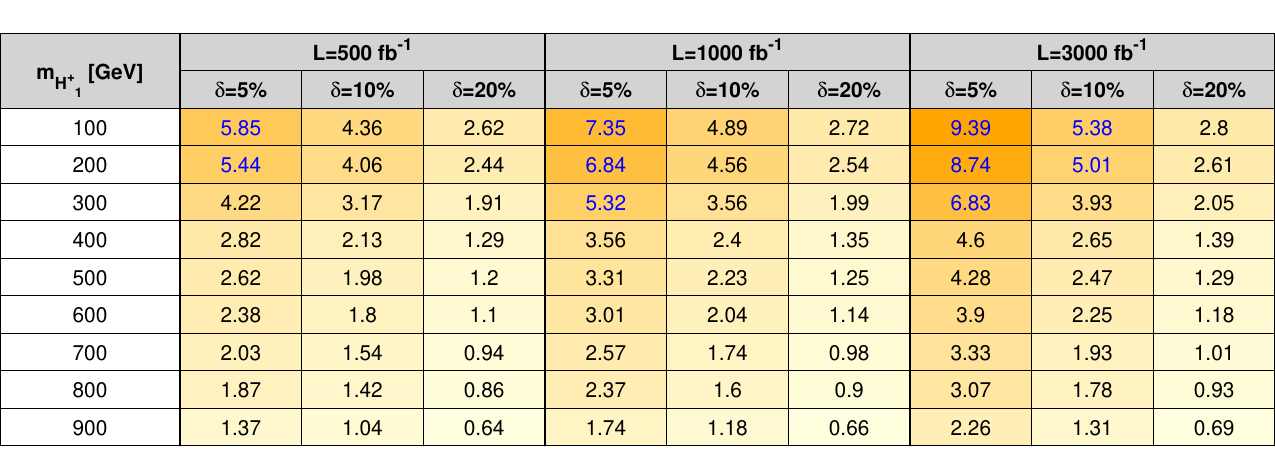}	
	\caption{The significance for the process $\mu^+ \mu^- \to H_1^\pm H_2^\mp \to \tau^+ \nu_{\tau} \tau^- \nu_{\tau}$ is presented as a function of the charged Higgs mass at a 3 TeV muon collider, considering integrated luminosities of 500 fb$^{-1}$, 1000 fb$^{-1}$, and 3000 fb$^{-1}$ and varying systematic uncertainties $\delta$, using our benchmark points presented in Table~\ref{Hp1Hm2_BPs_table}.}
	\label{SigHp1Hm2_table}
\end{table*}

\begin{table*}[t]
	\scriptsize
	\setlength{\tabcolsep}{1.6pt}
	\renewcommand{\arraystretch}{1.3}
	\centering
	\begin{tabular}{|c|cccccccccccccccccc|}
		\hline \hline
		& $m_{h_1}$ & $m_{h_2}$ & $\lambda_1$ & $\lambda_3$ & $\lambda_4$ & $\lambda_6$ & $\lambda_7$ & $\lambda_8$ & $\lambda_9$ & $\bar{\lambda}_8$ & $\bar{\lambda}_9$ & $\tan\beta$ & $\alpha_1$ & $\alpha_2$ & $\alpha_3$ & $v_t$ & $\mu_1$ \\
		\hline
		BP1'' & 125.09 & 128.72 & 0.48 & 0.0562 & 2.05 & 1.26 & 0.0184 & -0.659 & 0.404 & 0.829 & 2.38 & 16.23 & 1.52 & -0.0397 & -0.597 & 1.24 & -13.91 \\
		BP2'' & 125.09 & 209.91 & 3.32 & 1.09 & 1.59 & 0.77 & 0.267 & 0.842 & 0.135 & 0.406 & 1.00 & 11.47 & -1.49 & 0.0418 & 0.797 & 1.12 & -27.80 \\
		BP3'' & 125.09 & 298.86 & 4.56 & 0.558 & 2.42 & 0.309 & 0.909 & 0.120 & 0.0615 & 0.733 & 2.99 & 13.62 & 1.51 & -1.06e-3 & -0.615 & 0.0758 & 53.69 \\
		BP4'' & 125.09 & 435.25 & 1.25 & 5.03 & 2.44 & 2.25 & 0.936 & 0.622 & 1.13 & 1.71 & -0.254 & 15.26 & -1.51 & 2.80e-3 & -1.49 & 1.02 & -27.65 \\
		BP5'' & 125.09 & 422.08 & 0.236 & 3.69 & -1.78 & 2.48 & 1.83 & -0.156 & -0.620 & 1.63 & -1.39 & 20.93 & 1.53 & 1.48e-3 & 0.671 & 0.732 & 75.72 \\
		BP6'' & 125.09 & 577.95 & 0.501 & 3.19 & -1.91 & 3.23 & 1.88 & 0.179 & 3.18 & 0.295 & 0.808 & 14.29 & 1.51 & -7.05e-3 & -0.693 & 1.12 & 44.29 \\
		BP7'' & 125.09 & 697.34 & 0.328 & 0.469 & 1.85 & 2.03 & 1.44 & 1.32 & 2.35 & 0.766 & 1.50 & 13.28 & 1.50 & 1.00e-3 & 0.835 & 0.159 & 23.23 \\
		BP8'' & 125.09 & 798.02 & 0.177 & 1.79 & 0.650 & 2.21 & 1.02 & -1.39 & 2.10 & 2.09 & -1.19 & 13.70 & 1.51 & -2.06e-3 & -0.732 & 0.295 & 73.15 \\
		BP9'' & 125.09 & 869.58 & 0.827 & 2.78 & -0.891 & 5.17 & 3.77 & -1.31 & -1.98 & 0.673 & 0.738 & 13.02 & 1.50 & 5.28e-4 & 0.665 & 0.0477 & 51.94 \\
		\hline \hline
	\end{tabular}
	\caption{Benchmark points used in Table~\ref{SigHp1Hm2_table}.}
	\label{Hp1Hm2_BPs_table}
\end{table*}

\begin{table*}[t]
	\scriptsize
	\setlength{\tabcolsep}{2.2pt}
	\renewcommand{\arraystretch}{1.3}
	\centering
	\begin{tabular}{|c|cccccccccccccccccc|}
		\hline \hline
		& $m_{h_1}$ & $m_{h_2}$ & $\lambda_1$ & $\lambda_3$ & $\lambda_4$ & $\lambda_6$ & $\lambda_7$ & $\lambda_8$ & $\lambda_9$ & $\bar{\lambda}_8$ & $\bar{\lambda}_9$ & $\tan\beta$ & $\alpha_1$ & $\alpha_2$ & $\alpha_3$ & $v_t$ & $\mu_1$ \\
		\hline
		BP1''' & 125.09 & 433.08 & 5.26 & -0.28 & 2.97 & 2.87 & 0.38 & -1.49 & -0.27 & 3.40 & -1.26 & 117.51 & -1.56 & -3.84e-4 & -0.0779 & 0.0772 & -79.66 \\
		BP2''' & 125.09 & 271.23 & 0.19 & 1.04 & 0.69 & 6.89 & 3.50 & -3.90 & 2.57 & 0.47 & -0.15 & 119.38 & -1.56 & 4.54e-5 & -2.62e-3 & 0.0553 & -85.48 \\
		BP3''' & 125.09 & 545.02 & 0.24 & 1.83 & 3.67 & 1.13 & 3.66 & 0.58 & -2.60 & 1.84 & -0.46 & 118.47 & -1.56 & -3.75e-4 & -0.0397 & 0.159 & 69.36 \\
		BP4''' & 125.09 & 395.63 & 0.81 & 4.32 & -0.82 & 2.47 & 3.27 & -0.33 & 3.38 & -0.27 & 1.30 & 118.61 & -1.56 & 3.87e-5 & -0.0326 & 0.249 & -62.74 \\
		BP5''' & 125.09 & 500.69 & 0.62 & 7.75 & -1.60 & 3.76 & 1.65 & 1.47 & 1.86 & -0.61 & 1.29 & 112.07 & -1.56 & 2.37e-4 & 0.0176 & 0.0181 & -29.40 \\
		BP6''' & 125.09 & 501.97 & 0.095 & 3.54 & -1.19 & 3.38 & 1.81 & 1.81 & -0.29 & 0.97 & -0.50 & 117.09 & 1.56 & -1.31e-4 & 0.998 & 0.309 & -75.52 \\
		BP7''' & 125.09 & 551.69 & 0.25 & 6.56 & -3.44 & 3.05 & 4.49 & -1.07 & -0.12 & 1.81 & -1.11 & 118.12 & 1.56 & -2.22e-4 & 0.0178 & 0.200 & 16.78 \\
		BP8''' & 125.09 & 828.65 & 0.48 & 0.63 & 1.10 & 4.96 & 4.10 & -1.80 & -0.63 & 3.79 & -2.69 & 106.03 & 1.56 & -8.01e-5 & 0.0782 & 0.105 & 21.19 \\
		BP9''' & 125.09 & 714.86 & 0.47 & 9.47 & -3.98 & 4.45 & 3.15 & -1.37 & 1.40 & -0.16 & 0.92 & 114.24 & 1.56 & -1.82e-4 & 0.99 & 0.262 & -0.30 \\
		\hline \hline
	\end{tabular}
	\caption{Benchmark points used in Table~\ref{SigHp1W_table}.}
	\label{Hp1W_BPs_table}
\end{table*}

{\color{blue}\[ \mu^{+} \mu^{-} \rightarrow H_1^{\pm} H_2^{\mp} \rightarrow  \tau^+ \nu \tau^- \nu \]}

A detailed signal-to-background analysis of the process $\mu^{+} \mu^{-} \rightarrow H_1^{\pm} H_2^{\mp} \rightarrow \tau^+ \nu \tau^- \nu$ has been performed based on the cross section results presented in Fig.\ref{results_Hp1Hm2}. The outcome of this analysis indicates that the statistical significance remains extremely low, implying that an integrated luminosity exceeding approximately 16,000 fb$^{-1}$ would be necessary to achieve a a $5\sigma$ significance. This result can be primarily attributed to the fact that the production cross section for this process is considerably suppressed. Such suppression arises from imposing the conditions $\alpha_{2,3} \approx 0$, which are necessary to remain consistent with the approximations adopted in the computation of the electroweak oblique parameters S and T, as discussed in Refs.\cite{Ouazghour:2018mld,Ouazghour:2023eqr}.

However, as shown in Fig.~\ref{results_Hp1Hm2_NO_STU}, if we relax these constraints and instead consider a wider parameter range for $\alpha_{2,3}$, namely $-\pi/2 < \alpha_{2,3} < \pi/2$, the situation changes significantly. In this case, we can achieve larger values for the different couplings involved, which in turn enhances the production cross section, specifically in the low mass regions of $m_{H_1^{\pm}}$. Consequently, the cross section can reach values up to approximately 0.62 fb, suggesting that a bigger significance can be achieved.

Similar to the pair production process the events of the signal and backgrounds at parton level are required to pass through the basic cuts as follows : 
\begin{equation} \nonumber
	p_{T}^{j} > 20 \hspace*{4mm},\hspace*{4mm} \hspace*{4mm}  |\eta^{j}| < 2.5 ;
\end{equation}
Figure \ref{dist_Hp1Hm2} presents the kinematic distributions of the signal and background processes at a 3 TeV muon collider. The left panel shows the pseudorapidity of the leading tau, $\eta(\tau_1)$, and the right panel displays the transverse mass, $M_{T}\ [\tau_{1}]\ $. To enhance signal significance, we employ a cut-flow strategy based on these kinematic distributions, as detailed in Table \ref{cutft_HP1Hm2}. The first step in this strategy involves constraining the number of b-quarks and jets, imposing the conditions $N(b) \leq 1$ and $N(j) \leq 1$. This constraint plays a huge role in distinguishing the signal from the background. After applying this cut, only 21.6$\%$ of $Zjj$, 39.4$\%$ of $Wjj$ and 44.9$\%$ of $t\bar t$ background events remain, while nearly 100$\%$ of the signal events are preserved. 

Next, we impose the cut $-0.8 < \eta [\tau_1] < 0.8$, which effectively eliminates most of the $ZZ$, $Zjj$, and $Wjj$ background events, while retaining approximately 70$\%$ of $t\bar{t}$ and 69.1 $\%$of the signal events. Finally, an additional cut of $M_{T}\ [\tau_{1}] > 200$ GeV, which primarily suppress $Wjj$ backgrounds while preserving around 78.7$\%$ of the signal events.

Exactly as was done for the pair production process discussed previously, we perform an analysis covering a mass range from 100 GeV to 900 GeV. Following the behavior of the kinematic distributions for the benchmark point (BP2), we present the discovery significance ($\mathcal{Z}_\mathrm{disc}$) for each mass point at a center-of-mass energy of $\sqrt{s} = 3$ TeV, considering integrated luminosities of 500 fb$^{-1}$, 1000 fb$^{-1}$, and 3000 fb$^{-1}$, as well as systematic uncertainties ($\delta$) of 5$\%$, 10$\%$, and 20$\%$. All evaluated points satisfy both theoretical and experimental constraints.

The results show that at the lower bound of projected luminosity, $\mathcal{L} = 500$ fb$^{-1}$, the significance exceeds the $5\sigma$ discovery threshold ($\mathcal{Z}_\mathrm{disc} \geq 5$) for $m_{H_1^{\pm}}=100, 200$ GeV when $\delta = 5\%$. For the other considered systematic uncertainties ($\delta = 10\%$ and $20\%$), the significance remains below the discovery threshold ($\mathcal{Z}_\mathrm{disc} \leq 5$).

For an integrated luminosity of $\mathcal{L} = 1000$ fb$^{-1}$ and $\delta = 5\%$, the significance exceeds the discovery threshold ($\mathcal{Z}_\mathrm{disc} \geq 5$) for $m_{H_1^{\pm}}=100, 200, 300$ GeV. However, for $\delta = 10\%$ and $20\%$, the significance drops below the discovery threshold. Finally, for $\mathcal{L} = 3000$ fb$^{-1}$, the significance remains above the discovery threshold ($\mathcal{Z}_\mathrm{disc} \geq 5$) even for $\delta = 10\%$ for $m_{H_1^{\pm}}=100, 200$ GeV.

			\begin{table*}[t]
	\centering
	\renewcommand{\arraystretch}{1.3}
	\setlength{\tabcolsep}{45pt}
	\begin{adjustbox}{max width=\textwidth}
		\begin{tabular}{l c} 
			\hline \hline
			\textbf{Cuts}  & \textbf{Definition} \\  
			\hline \hline	
			\textbf{Trigger} & $N(b) \leq 1$\quad and \quad $N(j) \leq 1$ \\  	
			\hline
			\textbf{Cut-1} & $-0.8 < \eta [\tau_{1}] < 0.8$  \\  
			\hline
			\textbf{Cut-2} & $M_T[\tau_{1}] > 200$ GeV \\
			\hline \hline
		\end{tabular}
	\end{adjustbox}		
	\caption{Selection criteria applied in the signal-background analysis of the process 
		$\mu^+ \mu^- \to H_1^+ H_2^- \to \tau^+ \nu \tau^- \nu$ 
		at $\sqrt{s} = 3$ TeV.}
	\label{cutft_HP1Hm2}
\end{table*}
			\begin{figure*}[!htb]
	\centering	
	\includegraphics[scale=0.35]{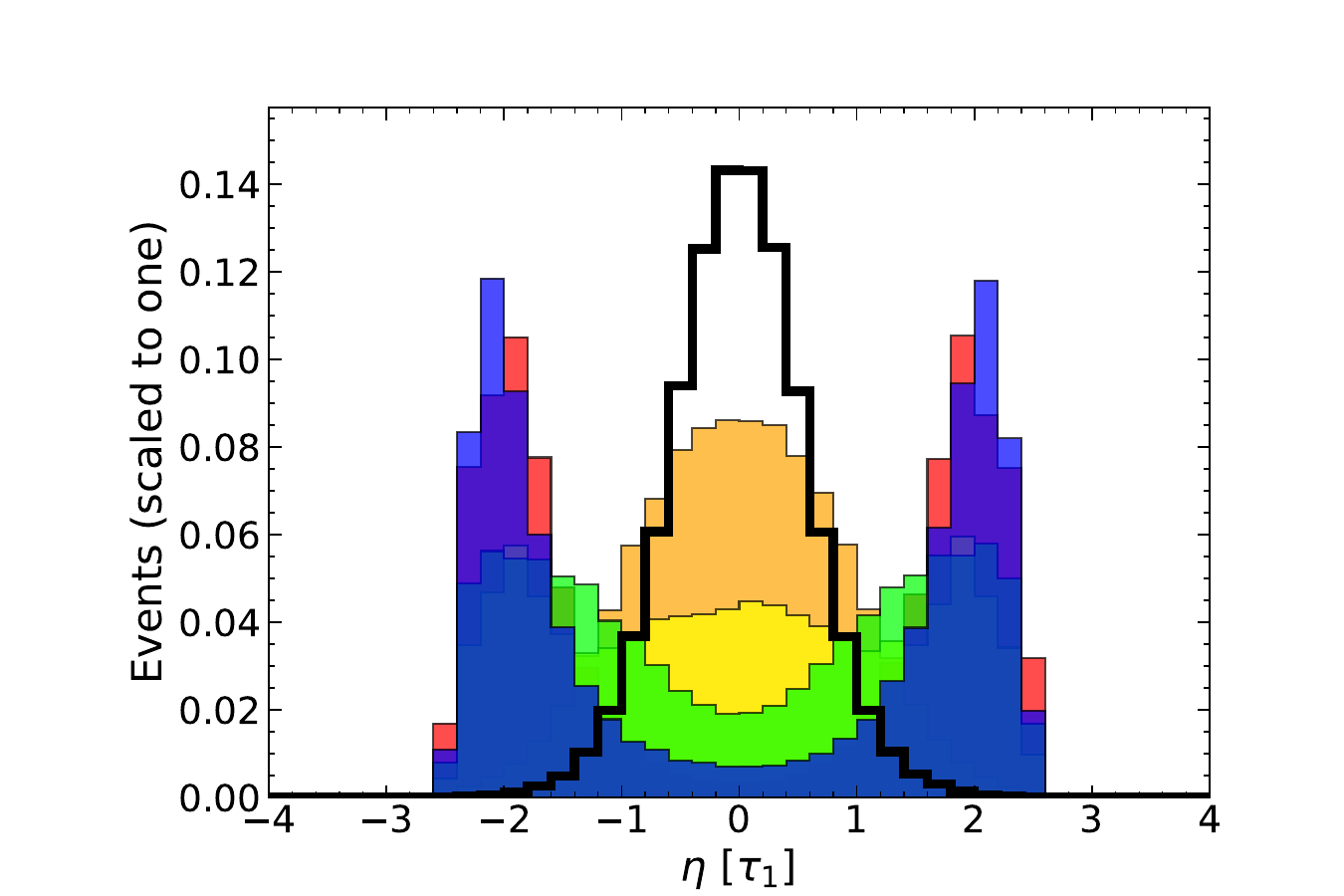}
	\includegraphics[scale=0.35]{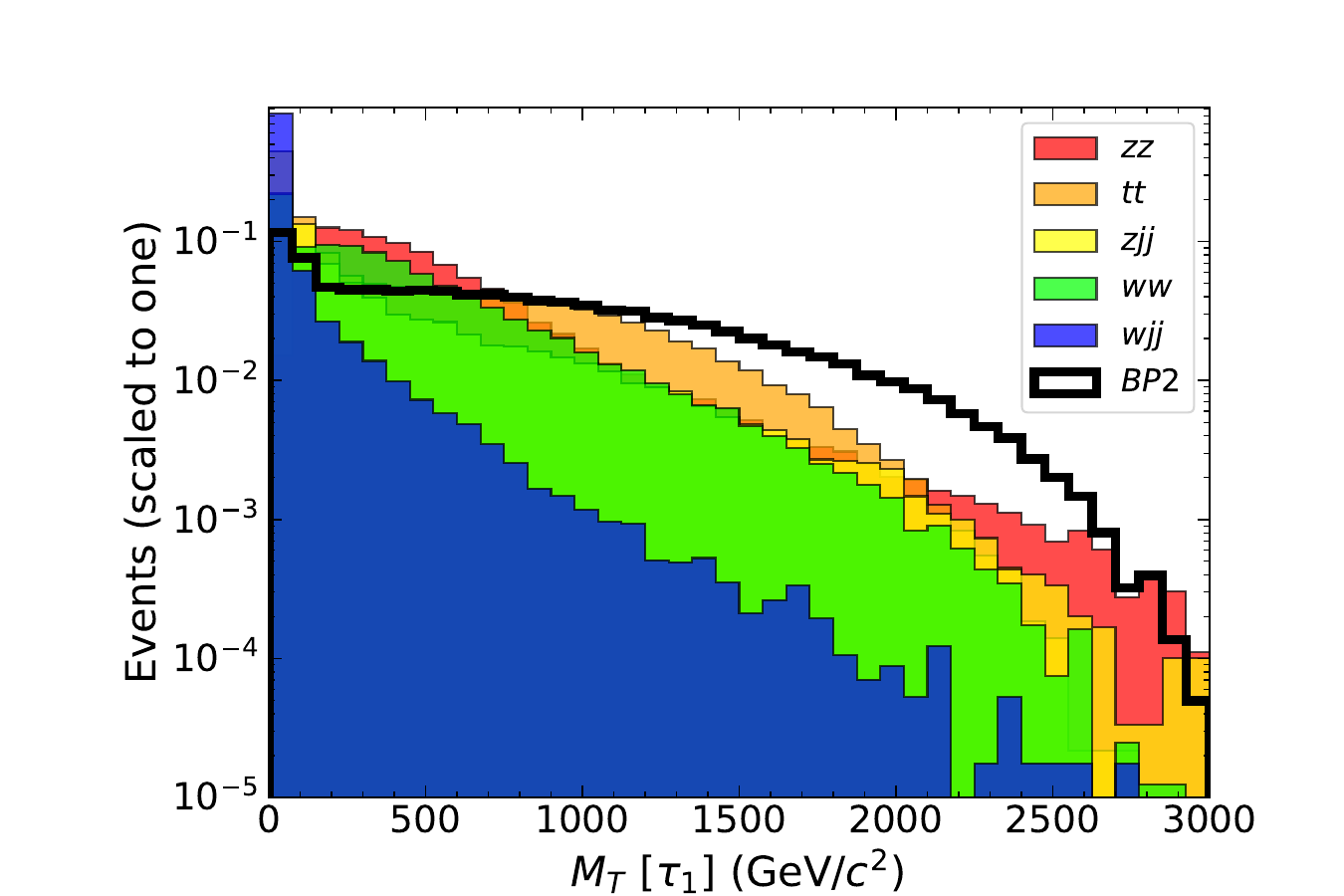}
	\caption{Normalized kinematic distributions of the signal and backgrounds: the pseudorapidity $\eta(\tau_1)$ (left panel) and the transverse mass $M_{T}\ [\tau_{1}]\ (\mathrm{GeV}/c^{2})$ (right panel) at $\sqrt{s}=$3 TeV muon collider.} \label{dist_Hp1Hm2}
\end{figure*}

		\begin{table*}[!tb]
	
	\setlength{\tabcolsep}{7pt}
	\renewcommand{\arraystretch}{1.2}
	\centering
	\begin{tabular}{l c c c c c c}
		\hline\hline
		\multirow{2}{*}{\textbf{Cuts}} & \textbf{Signal} & \multicolumn{5}{c}{\textbf{Backgrounds}} \\ 
		\cline{2-2} \cline{3-7} 
		& BP2 & $WW$ & $ZZ$ & $t\bar t$ & $Wjj$ & $Z/\gamma jj$ \\
		\hline\hline
        Basic cut  & 0.4580 & 1.54 & 0.0147 & 0.21 & 7.04 & 0.74 \\
        Tagger  & 0.4535 & 1.508 & 0.01465 & 0.0901 & 2.76 & 0.16 \\
        Cut-1  & 0.313 & 0.234 & 0.000417 & 0.0639 & 0.0564 & 0.0133 \\
        Cut-2  & 0.246 & 0.166 & 0.000382 & 0.0517 & 0.0162 & 0.00949 \\
		\hline
		\textbf{Total efficiencies} & \textbf{53.8\%} & \textbf{12.4\%} & \textbf{2.8\%} & \textbf{31.3\%} & \textbf{0.38\%} & \textbf{1.6\%} \\
		\hline\hline
	\end{tabular}
	\caption{Cut flow of the cross sections (in fb) for the signal and SM backgrounds at $\sqrt{s} = 3$ TeV muon collider using the benchmark point (BP2).}
	\label{cut2_Hp1Hm2}
\end{table*}
{\color{blue}\[ \mu^{+} \mu^{-} \rightarrow H_1^{\pm} W^{\mp} \rightarrow  \tau^+ \nu \tau^- \nu \]}
Similar to the previous processes the events of the signal and backgrounds at parton level are required to pass through the basic cuts as follows : 
			\begin{equation} \nonumber
				p_{T}^{j} > 20 \hspace*{4mm},\hspace*{4mm} \hspace*{4mm}  |\eta^{j}| < 2.5 ;
			\end{equation}
Figure \ref{dist_Hp1W} presents the kinematic distributions of the signal and background processes at a 3 TeV muon collider. The left panel shows the pseudorapidity of the leading tau, $\eta(\tau_1)$, the middle panel shows the transverse mass $M_T[\tau_{1}]$ and the right panel displays the transverse energy, $E_{T}$. To enhance signal significance, we employ a cut-flow strategy based on these kinematic distributions, as detailed in Table \ref{cut2_Hp1W}. The first step in this strategy involves constraining the number of b-quarks and jets, imposing the conditions $N(b) \leq 1$ and $N(j) \leq 1$. This constraint plays a key role in distinguishing the signal from the background. After applying this cut, only 21.6$\%$ of $Zjj$, 39.4$\%$ of $Wjj$ and 44.9$\%$ of $t\bar t$ background events persist, while nearly 100$\%$ of the signal events are preserved. 
			
Next, we impose the cut $-1.6 < \eta [\tau_1] < 1.6$, which effectively eliminates most of the $ZZ$, $Zjj$, and $Wjj$ background events while retaining approximately 96$\%$ of $t\bar{t}$ and 61.3 $\%$of the signal events. Finally, an additional cut of $E_{T} < 600$ GeV, which primarily suppress $t\bar{t}$, $Zjj$, and $Wjj$ backgrounds while preserving around 70$\%$ of the signal events. Then an additional cut on the transverse mass $M_T[\tau_{1}] > 50$ GeV eliminates about 65$\%$ of $Wjj$ and 23$\%$ of $WW$, while keeping about 98$\%$ of the signal. The selection cuts imposed are listed in Table~\ref{tab:cuts}.
			\begin{figure*}[!htb]
				\centering	
				\includegraphics[scale=0.26]{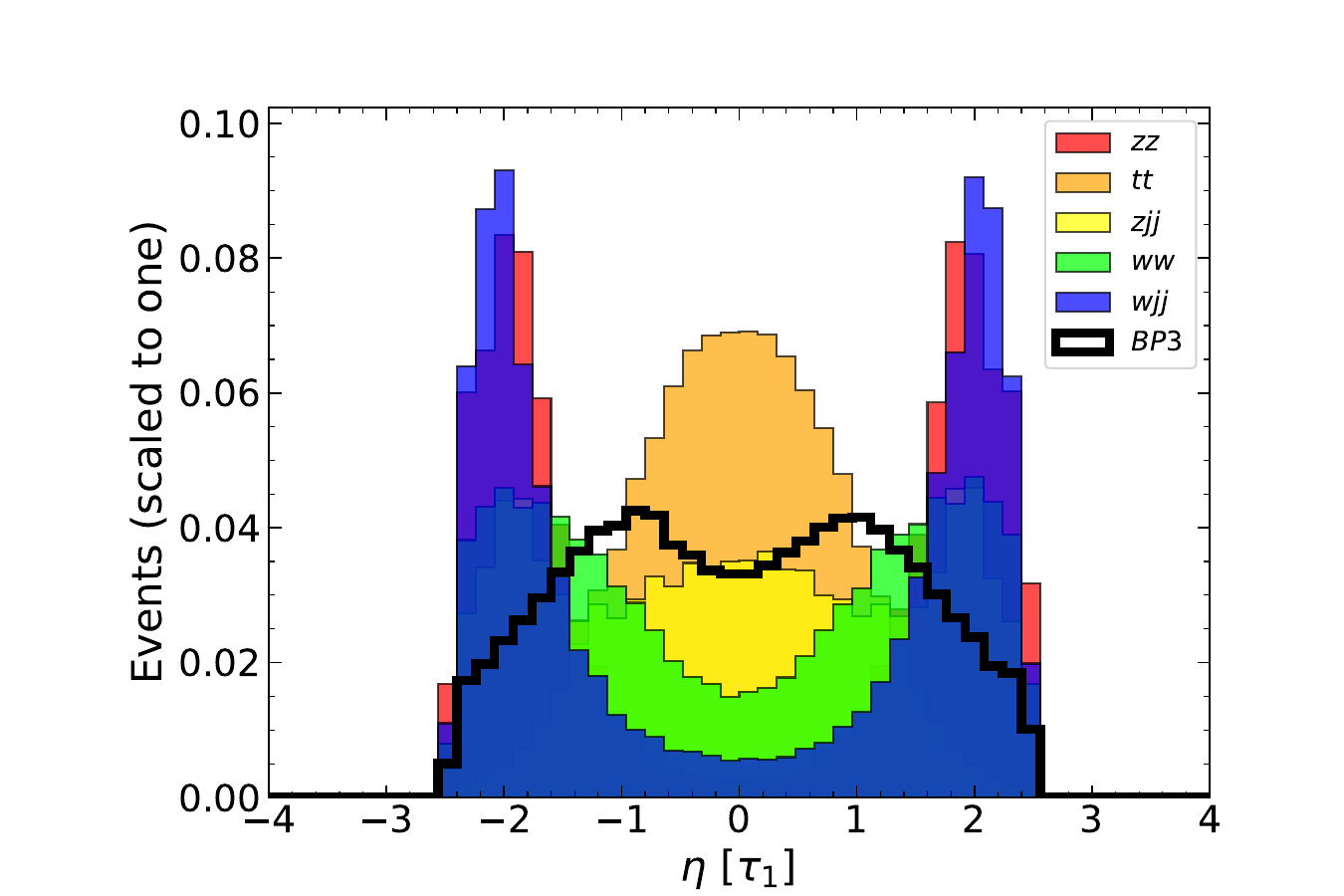}
				\includegraphics[scale=0.26]{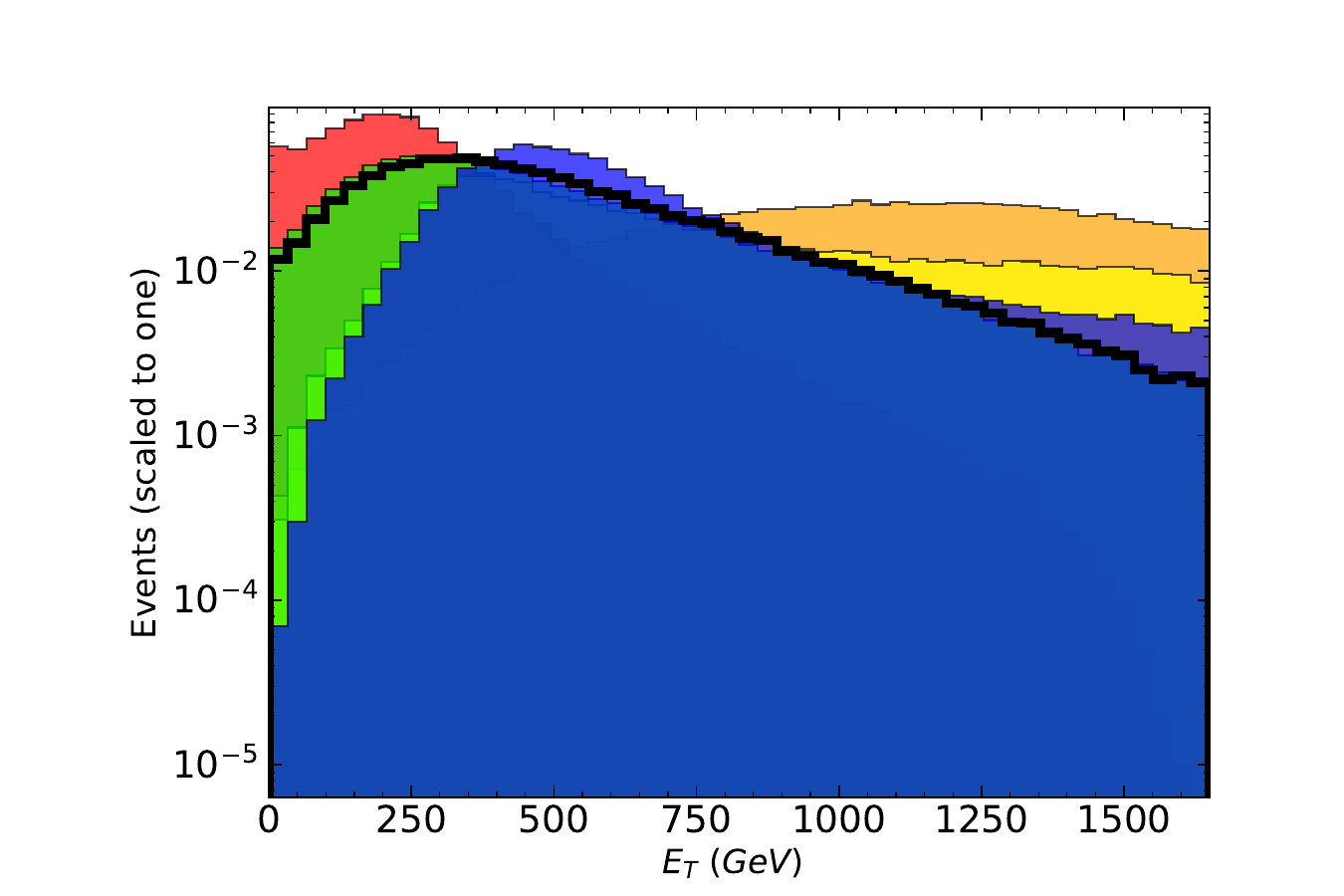}
		    	\includegraphics[scale=0.26]{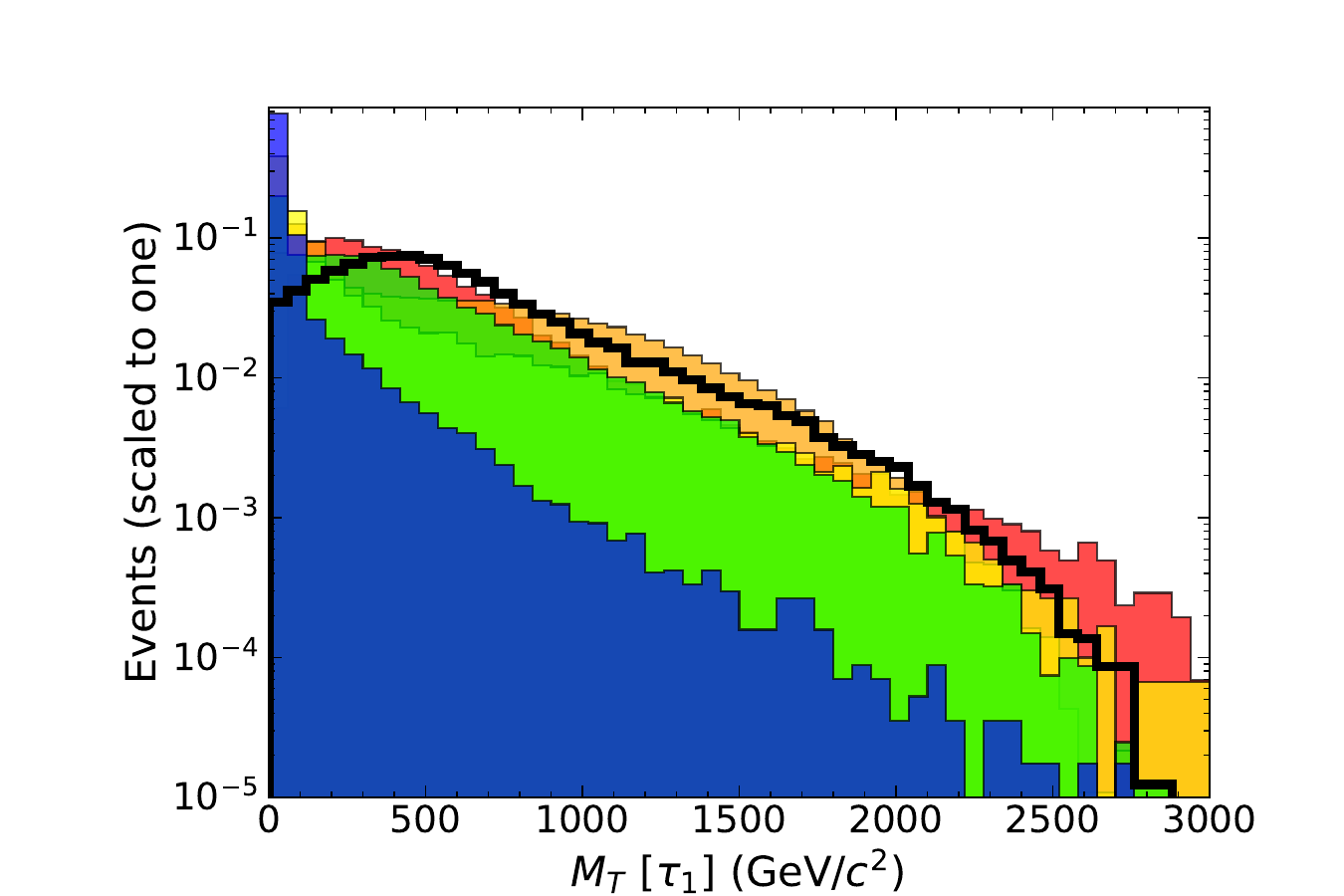}	\caption{Normalized kinematic distributions of the signal and backgrounds: the pseudorapidity $\eta(\tau_1)$ (left panel), the transverse energy $E_T$ (middle panel) and the transverse mass $M_T[\tau_{1}]$ (right panel) at $\sqrt{s}=$3 TeV muon collider.} \label{dist_Hp1W}
			\end{figure*}
			\begin{table*}[t]
				\centering
				\renewcommand{\arraystretch}{1.3}
				\setlength{\tabcolsep}{50pt}
				\begin{adjustbox}{max width=\textwidth}
					\begin{tabular}{l c} 
						\hline \hline
						\textbf{Cuts}  & \textbf{Definition} \\  
						\hline \hline
						\textbf{Trigger}  & $N(b) \leq 1$\quad and \quad $N(j) \leq 1$\\  
						\hline
						\textbf{Cut-1}  & $-1.6 < \eta [\tau_{1}] < 1.6$ \\  
						\hline
						\textbf{Cut-2}  & $\slashed{E_{T}} < 600$ GeV \\  
						\hline
						\textbf{Cut-3}  & $M_T[\tau_{1}] > 50$ GeV \\
						\hline \hline
					\end{tabular}
				\end{adjustbox}	
				\caption{The selection cuts applied in the signal-background analysis for the process $\mu^+ \mu^- \to H_1^\pm W^\mp \to \tau^+ \nu_{\tau} \tau^- \nu_{\tau}$ at $\sqrt{s} = 3$ TeV.}	
				\label{tab:cuts}
			\end{table*}
			\begin{table*}[!tb]

				\setlength{\tabcolsep}{7pt}
				\renewcommand{\arraystretch}{1.2}
				\centering
				\begin{tabular}{l c c c c c c}
					\hline\hline
					\multirow{2}{*}{\textbf{Cuts}} & \textbf{Signal} & \multicolumn{5}{c}{\textbf{Backgrounds}} \\ 
					\cline{2-2} \cline{3-7} 
					& BP3 & $WW$ & $ZZ$ & $t\bar t$ & $Wjj$ & $Z/\gamma jj$ \\
					\hline\hline
					Basic cut  & 0.4089 & 1.54 & 0.0147 & 0.2 & 7.04 & 0.74 \\
					Tagger  & 0.4045 & 1.508 & 0.01465 & 0.0901 & 2.76 & 0.16 \\
					Cut-1  & 0.248 & 0.667 & 0.00272 & 0.0868 & 0.449 & 0.0287 \\
					
					Cut-2  & 0.116 & 0.274 & 0.0019 & 0.00599 & 0.0222 & 0.00267 \\
					
					Cut-3  & 0.113 & 0.211 & 0.00197 & 0.00559 & 0.00778 & 0.002 \\
					\hline
					\textbf{Total efficiencies} & \textbf{30.6\%} & \textbf{13.7\%} & \textbf{13.4\%} & \textbf{2.79\%} & \textbf{0.11\%} & \textbf{0.26\%} \\
					\hline\hline
				\end{tabular}
				\caption{Cut flow of the cross sections (in fb) for the signal and SM backgrounds at $\sqrt{s} = 3$ TeV muon collider using the benchmark point (BP3).}
				\label{cut2_Hp1W}
			\end{table*}
\begin{table*}[!htb]
	\centering
	\includegraphics[scale=0.70]{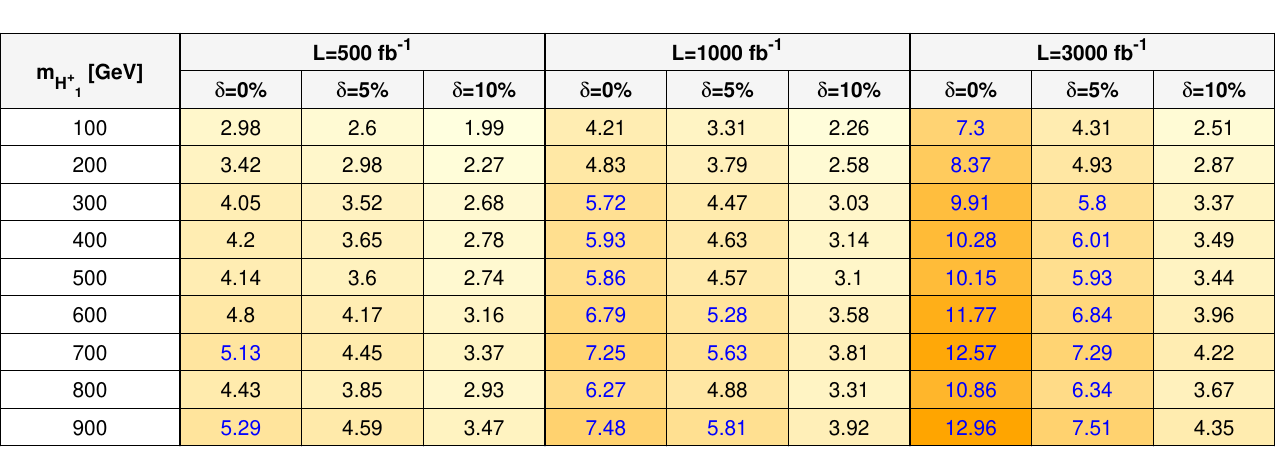}	
	\caption{The significance for the process $\mu^+ \mu^- \to H_1^\pm W^\mp \to \tau^+ \nu_{\tau} \tau^- \nu_{\tau}$ is presented as a function of the charged Higgs mass at a 3 TeV muon collider, considering integrated luminosities of 500 fb$^{-1}$, 1000 fb$^{-1}$, and 3000 fb$^{-1}$ and varying systematic uncertainties $\delta$, using our benchmark points presented in Table~\ref{Hp1W_BPs_table}.}
	\label{SigHp1W_table}
\end{table*}

Similarly to the previous processes, and following the behavior of the kinematic distributions for the benchmark point (BP3), we present the discovery significance ($\mathcal{Z}_\mathrm{disc}$) for different values of the charged Higgs mass at a center-of-mass energy of $\sqrt{s} = 3$ TeV, considering integrated luminosities of 500 fb$^{-1}$, 1000 fb$^{-1}$, and 3000 fb$^{-1}$, as well as systematic uncertainties ($\delta$) of 5$\%$ and 10$\%$. All evaluated points satisfy both theoretical and experimental constraints.

The results show that at the lowest considered luminosity, $\mathcal{L} = 500$ fb$^{-1}$, the significance exceeds the $5\sigma$ discovery threshold ($\mathcal{Z}_\mathrm{disc} \geq 5$) for $m_{H_1^{\pm}}=700, 900$ GeV when $\delta = 0\%$. For the other systematic uncertainties ($\delta = 5\%$ and $10\%$), the significance remains below the discovery threshold for all masses.

For an integrated luminosity of $\mathcal{L} = 1000$ fb$^{-1}$ and $\delta = 0\%$, the significance exceeds the discovery threshold for $m_{H_1^{\pm}}=300, 400, 500, 600, 700, 800, 900$ GeV. However, for $\delta = 5\%$, the significance drops significantly, with only $m_{H_1^{\pm}}=600, 700, 900$ remaining above the discovery threshold. Finally, for $\mathcal{L} = 3000$ fb$^{-1}$, the significance stays above the discovery threshold for all the masses considered, except for $m_{H_1^{\pm}}=100, 200$ GeV, which drop below the threshold when $\delta = 5\%$.

Overall, these results demonstrate that a future muon collider operating at $\sqrt{s} = 3$ TeV, offers significant discovery potential for charged Higgs bosons in the studied decay channel. These findings underscore the strong physics reach of such future collider facilities.
			
\section{CONCLUSION}
\vspace{6pt} 					
In this work, we have explored the phenomenology of the charged Higgs boson $H_1^{\pm}$ within the framework of the Two-Higgs-Doublet Type-II Seesaw Model (2HDMcT Type-III) at future $\mu^+ \mu^-$ colliders. Our investigation focused on $2 \to 2$ processes, such as $\mu^+ \mu^- \to H^+_1 S^-$ ($S^- = H^-_1, H^-_2$) and $\mu^+ \mu^- \to H^+_1 W^-$, incorporating both theoretical and experimental constraints.

The charged Higgs boson can be probed via $\mu^+ \mu^- \to H_1^+ H_1^-$ only for $m_{H_1^\pm} < \sqrt{s}/2$. Conversely, the process $\mu^+ \mu^- \to W^\pm H_1^\mp$ allows access to the mass region $\sqrt{s}/2 \leq m_{H_1^\pm} \leq \sqrt{s}-m_W$, which is kinematically inaccessible to charged Higgs pair production.

We demonstrated that the cross section for $\mu^+ \mu^- \to H_1^+ H_1^-$ can reach, and even exceed, that of the corresponding $e^+ e^-$ process, $\sigma(e^+ e^- \to H_1^+ H_1^-)\approx\sigma(\mu^+ \mu^- \to \gamma,Z \to H_1^+ H_1^-)$. Meanwhile, $\sigma(\mu^+ \mu^- \to H_1^+ H_2^-)$ is approximately eighteen times smaller than that of charged Higgs pair production, however, upon relaxing the constraints from EWPO and keeping all the others, the cross section can be enhanced up to 0.62 fb. Additionally, we found that the cross section for $\mu^+ \mu^- \to H^+_1 W^-$ can be substantially enhanced, particularly due to the large $\tan\beta$ amplification characteristic of Type-III scenarios.

Furthermore, we conducted a signal-background analysis and determined the discovery ($5\sigma$) regions at a 3 TeV muon collider for the $\mu^+ \mu^- \to H_1^+ H_1^-$, $\mu^+ \mu^- \to H_1^+ H_2^-$, and $\mu^+ \mu^- \to W^\pm H_1^\mp$ processes. 

The findings underscore the strong physics reach of future muon collider facilities and highlight the enhanced production prospects for charged Higgs bosons within this theoretical framework, offering valuable insights into their potential discovery.
			\vspace{6pt}
			\label{conlusion}
						

			\vspace{3pt}
			\begin{widetext}
			\section*{Apendix : Coupling Definitions}
			\vspace{1pt}
			\small
			\begin{eqnarray}
				\bar{\lambda}_{h_1H^\pm_1H^\pm_1}&=&\frac{-1}{2}\left(2 \mathcal{C}_{21}^2 \left(\lambda _6 \mathcal{E}_{13} v_{\Delta }+\lambda _1 v_1 \mathcal{E}_{11}+\lambda _3 v_2 \mathcal{E}_{12}\right)\right.\nonumber\\
				&&+2 \mathcal{C}_{22}^2 \left(\lambda _7 \mathcal{E}_{13} v_{\Delta }+\lambda _2 v_2 \mathcal{E}_{12}+\lambda _3 v_1 \mathcal{E}_{11}\right)\nonumber\\
				&&+\mathcal{C}_{23}^2 \left(2 \mathcal{E}_{13} (2\overline{\lambda _{8}}+\overline{\lambda _{9}}) v_{\Delta }+\left(2 \lambda _6+\lambda _8\right) v_1 \mathcal{E}_{11}+\left(2 \lambda _7+\lambda _9\right) v_2 \mathcal{E}_{12}\right)\nonumber\\
				&&+\mathcal{C}_{22} \mathcal{C}_{23} \left(\sqrt{2} \lambda _9 \mathcal{E}_{12} v_{\Delta }+\sqrt{2} \lambda _9 v_2 \mathcal{E}_{13}-4 \mu _2 \mathcal{E}_{12}-2 \mu _3 \mathcal{E}_{11}\right)\nonumber\\
				&&+\mathcal{C}_{21} \left(\right.\mathcal{C}_{23} \left(\sqrt{2} \lambda _8 \mathcal{E}_{11} v_{\Delta }+\sqrt{2} \lambda _8 v_1 \mathcal{E}_{13}-4 \mu _1 \mathcal{E}_{11}-2 \mu _3 \mathcal{E}_{12}\right)\nonumber\\
				&&\left.+2 \mathcal{C}_{22} \left(\lambda _4+\lambda _5\right) \left(v_2 \mathcal{E}_{11}+v_1 \mathcal{E}_{12}\right)\left.\right)\right)
				\label{3.3}
			\end{eqnarray}
			\vspace{3pt}
			\begin{eqnarray}
				\bar{\lambda}_{h_2H^\pm_1H^\pm_1}&&-\frac{1}{2} \left( 2 \mathcal{C}_{21}^2 \left( \mathcal{E}_{21} \lambda_1 v_1 + \mathcal{E}_{22} \lambda_3 v_2 + \mathcal{E}_{23} \lambda_6 v_t \right) \right. \nonumber\\
				&&+ 2 \mathcal{C}_{22}^2 \left( \mathcal{E}_{21} \lambda_3 v_1 + \mathcal{E}_{22} \lambda_2 v_2 + \mathcal{E}_{23} \lambda_7 v_t \right) \nonumber\\
				&&+ \mathcal{C}_{22} \mathcal{C}_{23} \left( -4 \mathcal{E}_{22} \mu_{2} - 2 \mathcal{E}_{21} \mu_{3} + \sqrt{2} \mathcal{E}_{23} \lambda_9 v_2 + \sqrt{2} \mathcal{E}_{22} \lambda_9 v_t \right) \nonumber\\
				&&+ \mathcal{C}_{23}^2 \left( \mathcal{E}_{21} \left( 2 \lambda_6 + \lambda_8 \right) v_1 + \mathcal{E}_{22} \left( 2 \lambda_7 + \lambda_9 \right) v_2 + 4 \mathcal{E}_{23} \left( \overline{\lambda _{8}} + \overline{\lambda _{9}} \right) v_t \right) \nonumber\\
				&&+ \mathcal{C}_{21} \left( 2 \mathcal{C}_{22} \left( \lambda_4 + \lambda_5 \right) \left( \mathcal{E}_{22} v_1 + \mathcal{E}_{21} v_2 \right) \right. \nonumber\\
				&&+ \mathcal{C}_{23} \left( -4 \mathcal{E}_{21} \mu_{1} - 2 \mathcal{E}_{22} \mu_{3} + \sqrt{2} \mathcal{E}_{23} \lambda_8 v_1 + \sqrt{2} \mathcal{E}_{21} \lambda_8 v_t \right) \left. \right) \left. \right)
			\end{eqnarray}
			\begin{eqnarray}
				\bar{\lambda}_{h_3H^\pm_1H^\pm_1}&&-\frac{1}{2} \left( 2 \mathcal{C}_{21}^2 \left( \mathcal{E}_{31} \lambda_1 v_1 + \mathcal{E}_{32} \lambda_3 v_2 + \mathcal{E}_{33} \lambda_6 v_t \right) \right. \nonumber\\
				&&+ 2 \mathcal{C}_{22}^2 \left( \mathcal{E}_{31} \lambda_3 v_1 + \mathcal{E}_{32} \lambda_2 v_2 + \mathcal{E}_{33} \lambda_7 v_t \right) \nonumber\\
				&&+ \mathcal{C}_{22} \mathcal{C}_{23} \left( -4 \mathcal{E}_{32} \mu_{2} - 2 \mathcal{E}_{31} \mu_{3} + \sqrt{2} \mathcal{E}_{33} \lambda_9 v_2 + \sqrt{2} \mathcal{E}_{32} \lambda_9 v_t \right) \nonumber\\
				&&+ \mathcal{C}_{23}^2 \left( \mathcal{E}_{31} \left( 2 \lambda_6 + \lambda_8 \right) v_1 + \mathcal{E}_{32} \left( 2 \lambda_7 + \lambda_9 \right) v_2 + 4 \mathcal{E}_{33} \left( \overline{\lambda _{8}} + \overline{\lambda _{9}} \right) v_t \right) \nonumber\\
				&&+ \mathcal{C}_{21} \left( 2 \mathcal{C}_{22} \left( \lambda_4 + \lambda_5 \right) \left( \mathcal{E}_{32} v_1 + \mathcal{E}_{31} v_2 \right) \right. \nonumber\\
				&&+ \mathcal{C}_{23} \left( -4 \mathcal{E}_{31} \mu_{1} - 2 \mathcal{E}_{32} \mu_{3} + \sqrt{2} \mathcal{E}_{33} \lambda_8 v_1 + \sqrt{2} \mathcal{E}_{31} \lambda_8 v_t \right) \left. \right) \left. \right)
			\end{eqnarray}
				\begin{eqnarray}
					\bar{\lambda}_{h_1H^\pm_1H^\pm_2}&&\frac{1}{4} \left( - \mathcal{C}_{23} \left( -4 \mathcal{C}_{32} \mathcal{E}_{12} \mu_{2} - 2 \mathcal{C}_{32} \mathcal{E}_{11} \mu_{3} + \sqrt{2} \mathcal{C}_{32} \lambda_9 \left( \mathcal{E}_{13} v_2 + \mathcal{E}_{12} v_t \right) \right. \right. \nonumber\\
					&&+ \mathcal{C}_{31} \left( -4 \mathcal{E}_{11} \mu_{1} - 2 \mathcal{E}_{12} \mu_{3} + \sqrt{2} \mathcal{E}_{13} \lambda_8 v_1 + \sqrt{2} \mathcal{E}_{11} \lambda_8 v_t \right) \nonumber\\
					&&+ 2 \mathcal{C}_{33} \left( \mathcal{E}_{11} \left( 2 \lambda_6 + \lambda_8 \right) v_1 + \mathcal{E}_{12} \left( 2 \lambda_7 + \lambda_9 \right) v_2 + 4 \mathcal{E}_{13} \left( \overline{\lambda _{8}} + \overline{\lambda _{9}} \right) v_t \right) \left. \right) \nonumber\\
					&&- \mathcal{C}_{21} \left( \mathcal{C}_{33} \left( -4 \mathcal{E}_{11} \mu_{1} - 2 \mathcal{E}_{12} \mu_{3} + \sqrt{2} \mathcal{E}_{13} \lambda_8 v_1 + \sqrt{2} \mathcal{E}_{11} \lambda_8 v_t \right) \right. \nonumber\\
					&&+ 2 \left( \mathcal{C}_{32} \left( \lambda_4 + \lambda_5 \right) \left( \mathcal{E}_{12} v_1 + \mathcal{E}_{11} v_2 \right) + 2 \mathcal{C}_{31} \left( \mathcal{E}_{11} \lambda_1 v_1 + \mathcal{E}_{12} \lambda_3 v_2 + \mathcal{E}_{13} \lambda_6 v_t \right) \right) \left. \right) \nonumber\\
					&&- \mathcal{C}_{22} \left( \mathcal{C}_{33} \left( -4 \mathcal{E}_{12} \mu_{2} - 2 \mathcal{E}_{11} \mu_{3} + \sqrt{2} \mathcal{E}_{13} \lambda_9 v_2 + \sqrt{2} \mathcal{E}_{12} \lambda_9 v_t \right) \right. \nonumber\\
					&&+ 2 \left( \mathcal{C}_{31} \left( \lambda_4 + \lambda_5 \right) \left( \mathcal{E}_{12} v_1 + \mathcal{E}_{11} v_2 \right) + 2 \mathcal{C}_{32} \left( \mathcal{E}_{11} \lambda_3 v_1 + \mathcal{E}_{12} \lambda_2 v_2 + \mathcal{E}_{13} \lambda_7 v_t \right) \right) \left. \right) \left. \right)
				\end{eqnarray}
				\begin{eqnarray}
					\bar{\lambda}_{h_2H^\pm_1H^\pm_2}&&\frac{1}{4} \left( - \mathcal{C}_{23} \left( -4 \mathcal{C}_{32} \mathcal{E}_{22} \mu_{2} - 2 \mathcal{C}_{32} \mathcal{E}_{21} \mu_{3} + \sqrt{2} \mathcal{C}_{32} \lambda_9 \left( \mathcal{E}_{23} v_2 + \mathcal{E}_{22} v_t \right) \right. \right. \nonumber\\
					&&+ \mathcal{C}_{31} \left( -4 \mathcal{E}_{21} \mu_{1} - 2 \mathcal{E}_{22} \mu_{3} + \sqrt{2} \mathcal{E}_{23} \lambda_8 v_1 + \sqrt{2} \mathcal{E}_{21} \lambda_8 v_t \right) \nonumber\\
					&&+ 2 \mathcal{C}_{33} \left( \mathcal{E}_{21} \left( 2 \lambda_6 + \lambda_8 \right) v_1 + \mathcal{E}_{22} \left( 2 \lambda_7 + \lambda_9 \right) v_2 + 4 \mathcal{E}_{23} \left( \overline{\lambda _{8}} + \overline{\lambda _{9}} \right) v_t \right) \left. \right) \nonumber\\
					&&- \mathcal{C}_{21} \left( \mathcal{C}_{33} \left( -4 \mathcal{E}_{21} \mu_{1} - 2 \mathcal{E}_{22} \mu_{3} + \sqrt{2} \mathcal{E}_{23} \lambda_8 v_1 + \sqrt{2} \mathcal{E}_{21} \lambda_8 v_t \right) \right. \nonumber\\
					&&+ 2 \left( \mathcal{C}_{32} \left( \lambda_4 + \lambda_5 \right) \left( \mathcal{E}_{22} v_1 + \mathcal{E}_{21} v_2 \right) + 2 \mathcal{C}_{31} \left( \mathcal{E}_{21} \lambda_1 v_1 + \mathcal{E}_{22} \lambda_3 v_2 + \mathcal{E}_{23} \lambda_6 v_t \right) \right) \left. \right) \nonumber\\
					&&- \mathcal{C}_{22} \left( \mathcal{C}_{33} \left( -4 \mathcal{E}_{22} \mu_{2} - 2 \mathcal{E}_{21} \mu_{3} + \sqrt{2} \mathcal{E}_{23} \lambda_9 v_2 + \sqrt{2} \mathcal{E}_{22} \lambda_9 v_t \right) \right. \nonumber\\
					&&+ 2 \left( \mathcal{C}_{31} \left( \lambda_4 + \lambda_5 \right) \left( \mathcal{E}_{22} v_1 + \mathcal{E}_{21} v_2 \right) + 2 \mathcal{C}_{32} \left( \mathcal{E}_{21} \lambda_3 v_1 + \mathcal{E}_{22} \lambda_2 v_2 + \mathcal{E}_{23} \lambda_7 v_t \right) \right) \left. \right) \left. \right)
				\end{eqnarray}
				\begin{eqnarray}
					\bar{\lambda}_{h_3H^\pm_1H^\pm_2}&&\frac{1}{4} \left( - \mathcal{C}_{23} \left( -4 \mathcal{C}_{32} \mathcal{E}_{32} \mu_{2} - 2 \mathcal{C}_{32} \mathcal{E}_{31} \mu_{3} + \sqrt{2} \mathcal{C}_{32} \lambda_9 \left( \mathcal{E}_{33} v_2 + \mathcal{E}_{32} v_t \right) \right. \right. \nonumber\\
					&&+ \mathcal{C}_{31} \left( -4 \mathcal{E}_{31} \mu_{1} - 2 \mathcal{E}_{32} \mu_{3} + \sqrt{2} \mathcal{E}_{33} \lambda_8 v_1 + \sqrt{2} \mathcal{E}_{31} \lambda_8 v_t \right) \nonumber\\
					&&+ 2 \mathcal{C}_{33} \left( \mathcal{E}_{31} \left( 2 \lambda_6 + \lambda_8 \right) v_1 + \mathcal{E}_{32} \left( 2 \lambda_7 + \lambda_9 \right) v_2 + 4 \mathcal{E}_{33} \left( \overline{\lambda _{8}} + \overline{\lambda _{9}} \right) v_t \right) \left. \right) \nonumber\\
					&&- \mathcal{C}_{21} \left( \mathcal{C}_{33} \left( -4 \mathcal{E}_{31} \mu_{1} - 2 \mathcal{E}_{32} \mu_{3} + \sqrt{2} \mathcal{E}_{33} \lambda_8 v_1 + \sqrt{2} \mathcal{E}_{31} \lambda_8 v_t \right) \right. \nonumber\\
					&&+ 2 \left( \mathcal{C}_{32} \left( \lambda_4 + \lambda_5 \right) \left( \mathcal{E}_{32} v_1 + \mathcal{E}_{31} v_2 \right) + 2 \mathcal{C}_{31} \left( \mathcal{E}_{31} \lambda_1 v_1 + \mathcal{E}_{32} \lambda_3 v_2 + \mathcal{E}_{33} \lambda_6 v_t \right) \right) \left. \right) \nonumber\\
					&&- \mathcal{C}_{22} \left( \mathcal{C}_{33} \left( -4 \mathcal{E}_{32} \mu_{2} - 2 \mathcal{E}_{31} \mu_{3} + \sqrt{2} \mathcal{E}_{33} \lambda_9 v_2 + \sqrt{2} \mathcal{E}_{32} \lambda_9 v_t \right) \right. \nonumber\\
					&&+ 2 \left( \mathcal{C}_{31} \left( \lambda_4 + \lambda_5 \right) \left( \mathcal{E}_{32} v_1 + \mathcal{E}_{31} v_2 \right) + 2 \mathcal{C}_{32} \left( \mathcal{E}_{31} \lambda_3 v_1 + \mathcal{E}_{32} \lambda_2 v_2 + \mathcal{E}_{33} \lambda_7 v_t \right) \right) \left. \right) \left. \right)
				\end{eqnarray}

				\begin{eqnarray}	\lambda_{h_1H_1^{\pm}W^{\mp}}&=&(\mathcal{E}_{11}\mathcal{C}_{21}+\mathcal{E}_{12}\mathcal{C}_{22}+\sqrt{2}\mathcal{E}_{13}\mathcal{C}_{23})
					\nonumber  \\
					\lambda_{h_2H_1^{\pm}W^{\mp}}&=&(\mathcal{E}_{21}\mathcal{C}_{21}+\mathcal{E}_{22}\mathcal{C}_{22}+\sqrt{2}\mathcal{E}_{23}\mathcal{C}_{23})
					\nonumber  \\
					\lambda_{h_3H_1^{\pm}W^{\mp}}&=&(\mathcal{E}_{31}\mathcal{C}_{21}+\mathcal{E}_{32}\mathcal{C}_{22}+\sqrt{2}\mathcal{E}_{33}\mathcal{C}_{23})
					\nonumber  \\
					\lambda_{A_1H_1^{\pm}W^{\mp}}&=&(\mathcal{O}_{21}\mathcal{C}_{21}+\mathcal{O}_{22}\mathcal{C}_{22}+\sqrt{2}\mathcal{O}_{23}\mathcal{C}_{23})
					\nonumber  \\
					\lambda_{A_2H_1^{\pm}W^{\mp}}&=&(\mathcal{O}_{31}\mathcal{C}_{21}+\mathcal{O}_{32}\mathcal{C}_{22}+\sqrt{2}\mathcal{O}_{33}\mathcal{C}_{23})
					\nonumber  \\
					\lambda_{ZH_1^{\pm}W^{\mp}}&=&\frac{g^2}{2c_w}(\mathcal{C}_{12}s_w^2v_1+\mathcal{C}_{22}s_w^2v_2+\sqrt{2}\mathcal{C}_{23}(s_w^2+1)v_t)
					\nonumber  \\
					Y_{6}&=&\mathcal{C}_{21}/C_{\beta}
					\nonumber  \\
					Y_{7}&=&\mathcal{C}_{31}/C_{\beta}
				\end{eqnarray}
					\end{widetext}
			
\providecommand{\href}[2]{#2}\begingroup\raggedright\endgroup
\end{document}